\documentclass[
    reprint,
    amsmath,
    amssymb,showpacs,showkeys,
    aps,
    prd]{revtex4-2}
\usepackage{upgreek}
\usepackage{graphicx}
\usepackage{float} 
\usepackage{silence} 
\usepackage{bm}
\usepackage{slashed}
\usepackage{amsmath}
\usepackage{amssymb} 
\usepackage{lmodern} 
\usepackage{simplewick}
\usepackage{color}
\begin{document}
\title{Mixing angle of $K_1(1270/1400)$ and the $K\bar K_1(1400)$ molecular interpretation of $\eta_1(1855)$}
\author{Zheng-Shu Liu$^1$}
\author{Xu-Liang Chen$^1$}
\author{Ding-Kun Lian$^1$}
\author{Ning Li$^1$}
\email{lining59@mail.sysu.edu.cn}
\author{Wei Chen$^{1,\, 2}$}
  \email{chenwei29@mail.sysu.edu.cn}
\affiliation{$^1$School of Physics, Sun Yat-sen University, Guangzhou 510275, China \\
$^2$Southern Center for Nuclear-Science Theory (SCNT), Institute of Modern Physics, 
Chinese Academy of Sciences, Huizhou 516000, Guangdong Province, China}
\begin{abstract}
Due to the SU(3) symmetry breaking effect, the axial-vector kaons $K_1(1270)$ and $K_1(1400)$ are established to be mixtures of two P-wave $K_{1A}\left( {^3{P_1}} \right)$ and $K_{1B}\left( {^1{P_1}} \right)$ states. In QCD sum rules, we propose a new construction of the $K_1$ current operators and calculate the two-point correlation functions by including the next-to-leading order four-quark condensates. The mixing angle is determined as $\theta = \left( {46.95_{ - 0.23}^{ + 0.25}} \right)^\circ$ by reproducing the masses of $K_1(1270)$ and $K_1(1400)$. We further compose the $K\bar K_1\left( {1270} \right)$ and $K\bar K_1\left( {1400} \right)$ interpolating currents with exotic quantum numbers $J^{PC}=1^{-+}$ to investigate the possible molecular interpretation of the recently observed ${\eta _1}(1855)$ state. We calculate the correlation functions and perform the QCD sum rule analyses for these two molecular systems. However, the spectral functions are found to be negative in physical regions so that they are not able to provide reliable investigations of the $K\bar K_1$ molecular states. 
\end{abstract}
\pacs{12.39.Mk, 12.38.Lg, 14.40.Ev, 14.40.Rt}
\keywords{QCD sum rules, Mixing angle, Molecular states}
\maketitle
\section{Introduction}
Quantum chromodynamics (QCD) is the fundamental theory to study the hadron structure and hadron spectroscopy. In QCD, there exist various exotic hadron configurations beyond the conventional $q\bar q$ mesons and $qqq$ baryons~\cite{ParticleDataGroup:2022pth,Gell-Mann:1964ewy,Zweig:1964ruk}, such as the multiquark states, hybrid mesons, glueballs and so on
~\cite{Chen:2016qju,Guo:2017jvc,Liu:2019zoy,Brambilla:2019esw,Chen:2022asf}.

Recently, the BESIII Collaboration reported the existence of an isoscalar state $\eta_1(1855)$ in the $J/\psi \to \gamma\eta_1(1855)\to \gamma \eta \eta^\prime$ process with a statistical significance of more than $19\sigma$~\cite{BESIII:2022riz, BESIII:2022iwi}. The quantum numbers of $\eta_1(1855)$ have been determined as $I^GJ^{PC}= 0^+1^{-+}$, while its mass and decay width were measured as $M = 1855 \pm 9_{ - 1}^{ + 6}\;{\rm{MeV}}$ and $\Gamma  = 188 \pm 18_{ - 8}^{ + 3}\;{\rm{MeV}}$, respectively~\cite{BESIII:2022riz, BESIII:2022iwi}. Combined with the isovector state $\pi_1(1600)$, the observation of 
$\eta_1(1855)$ has been extensively considered to provide critical information about the $1^{-+}$ hybrid nonet~\cite{Qiu:2022ktc,Chen:2022qpd,Shastry:2022mhk,Wang:2022sib,Chen:2022isv,Swanson:2023zlm,Chen:2023ukh,Shastry:2023ths}.
On the other hand, the molecular interpretation of $\eta_1(1855)$ is also possible since its mass is just about 40 MeV below the $K\bar K_1(1400)$ threshold~\cite{Dong:2022cuw,Yang:2022rck}. Using the one boson exchange model, the authors of Ref.~\cite{Dong:2022cuw} investigated the attractive interaction and decay properties of $K\bar K_1(1400)$ molecule with $J^{PC}=1^{-+}$, which supported the molecular explanation of $\eta_1(1855)$. In Ref.~\cite{Yang:2022rck}, the radiative and strong decays of the S-wave $K\bar K_1(1400)$ molecular state were studied within the effective Lagrangian approach, and the result was confronted with the experimental data. Using the chiral unitary approach, the interactions between pseudoscalar and axial-vector mesons were studied to describe $\eta_1(1855)$ as a dynamically generated state~\cite{Yan:2023vbh}. 
Other different interpretations of $\eta_1(1855)$ can be found in Refs.~\cite{Su:2022eun,Wan:2022xkx}. Accordingly, the authors of Ref.~\cite{Huang:2022tpq} suggested detecting $\eta_1(1855)$ via photoproduction in order to distinguish its various interpretations.

In PDG~\cite{ParticleDataGroup:2022pth}, there are two physical axial-vector kaons $K_1(1270)$ and $K_1(1400)$ with quantum numbers $J^P=1^+$. Due to the SU(3) symmetry breaking, they are established to be mixtures of two P-wave states $K_{1A}\left( {^3{P_1}} \right)$ and $K_{1B}\left( {^1{P_1}} \right)$
     \begin{eqnarray}
       \begin{gathered}
         \left| {K_1(1270)} \right\rangle  = \sin \theta \left| {K_{1A}} \right\rangle  + \cos \theta \left| K_{1B} \right\rangle , \hfill \\
         \left| {K_1(1400)} \right\rangle  = \cos \theta \left| {K_{1A}} \right\rangle  - \sin \theta \left| K_{1B} \right\rangle , \hfill \label{mixingscheme}
       \end{gathered}
     \end{eqnarray}
where $\theta$ is the mixing angle. There are numerous studies of the value of $\theta$ in the literature. In Ref.~\cite{Suzuki:1993yc}, the mixing angle was determined to be $\theta  = {33^ \circ }$ or $57^ \circ$ by using the early experimental information on the masses and partial decay rates of $K_1(1270)$ and $K_1(1400)$.  
In Ref.~\cite{Blundell:1995au}, the authors estimated the mixing angle from the weak decays $\tau  \to K_1(1270/1400){\nu _\tau }$, concluding that $ - {30^ \circ } \leqslant \theta  \leqslant {50^ \circ }$. Their further analysis indicated that the most favored value is $\theta  \approx {45^ \circ }$. A nonrelativistic constituent quark model calculation provided a constraint on the mixing angle ${35^ \circ } < \theta  < {55^ \circ }$~\cite{Burakovsky:1997dd}, which is consistent with the value $\theta  = {39^ \circ } \pm {4^ \circ }$ obtained in the framework of QCD sum rules by assuming the orthogonality of the mass eigenstates~\cite{Dag:2012zz}. The studies of the charmed meson decays gave the preferred solution $\theta  \approx {-58^ \circ }$~\cite{Cheng:2003bn}. In Ref.~\cite{Hatanaka:2008xj}, the authors found that  $\theta  = {-(34\pm13)^ \circ }$ by comparing the light-cone sum rule calculation and the experimental data for $B\to K_1\gamma$ and $\tau  \to K_1(1270){\nu _\tau }$ decays. However, it is found to be $ \theta  \sim {33^ \circ }$ by studying the relations between $\theta$ and the mixing angle of $f_1(1285)-f_1(1420)$ $(h_1(1170)-h_1(1380))$~\cite{Cheng:2011pb,Cheng:2013cwa}. Recently, the mixing angle is determined to be $\theta  = {22^ \circ } \pm {7^ \circ }$ or $\theta  = {68^ \circ } \pm {7^ \circ }$ 
by calculating a $K_{1A}\to K_{1B}$ matrix element in the three-point QCD sum rules~\cite{Shi:2023kiy}. It is obviously that there is no consensus on the value of the mixing angle and the results from various approaches are still quite controversial. 

In this work, we shall present a different QCD sum rule calculation from that in Ref.~\cite{Dag:2012zz}. We propose a new construction for the current  operators of two axial-vector physical kaons and calculate the two-point correlation functions to extract the hadron masses. The mixing angle can be determined by reproducing the masses of $K_1(1270)$ and $K_1(1400)$. The construction of the $K_1$ operators is essential for studying the $K\bar K_1(1400)$ molecular interpretation of $\eta_1(1855)$. 

This work is organized as follows. In Sec. II, we introduce a new method to construct the current operators of $K_1(1270/1400)$ as the mixture of $K_{1A}\left( {^3{P_1}} \right)$ and $K_{1B}\left( {^1{P_1}} \right)$. We then determine the mixing angle by reproducing the masses of $K_1(1270)$ and $K_1(1400)$. In Sec. III, we construct the four-quark interpolating currents of $K\bar K_1(1270/1400)$ and investigate the possibility of the molecular interpretation of ${\eta _1}(1855)$. A brief summary is given in the last section. 

\section{QCD sum rules for the axial-vector $K_1$ mesons}
\subsection{Construction of the $K_1$ operators}
In general, the two P-wave states $K_{1A}\left( {^3{P_1}} \right)$ and $K_{1B}\left( {^1{P_1}} \right)$ can be coupled by the axial-vector and tensor currents respectively
\begin{eqnarray}
       J_A^\mu (x) = {\bar s}(x){\gamma ^\mu }{\gamma _5}q(x),\;J_B^{\mu \nu }(x) = {\bar s}(x){\sigma ^{\mu \nu }}{\gamma _5}q(x)\, ,\; \label{K1ABcurrents}
\end{eqnarray}
in which $s$ is the strange quark field and $q=u/d$ represents up or down quark field.
Considering the mixing scheme in Eq.~\eqref{mixingscheme}, we construct the interpolating currents for the mass eigenstates $K_1(1270)$ and $K_1(1400)$ 
as 
     \begin{eqnarray}
       \begin{gathered}
         J_{1270}^\mu (x) = J_A^\mu (x)\sin \theta  + J_B^\mu (x)\cos \theta ,\; \hfill \\
         J_{1400}^\mu (x) = J_A^\mu (x)\cos \theta  - J_B^\mu (x)\sin \theta ,\; \hfill
       \end{gathered}  \label{K1currents}
     \end{eqnarray}
where $J_B^\mu (x)$ is defined from the current $J_B^{\mu \nu }(x)$ in Eq.~\eqref{K1ABcurrents} and we shall give its specific form and discuss the detail later. 

Using the above interpolating currents, one can reproduce the masses of $K_1(1270)$ and $K_1(1400)$ within the QCD sum rules by calculating the two-point correlation function
\begin{eqnarray}
\Pi _X^{\mu \nu }\left( {{p^2}} \right) = {\rm{i}}\int {{{\rm{d}}^d}x} \;{{\rm{e}}^{{\rm{i}}p \cdot x}}\left\langle \Omega  \right|{\mathsf{T}}\left[ {J_X^\mu (x)J_X^{\nu \dag }(0)} \right]\left| \Omega  \right\rangle .
\end{eqnarray}
According to Eq.~\eqref{K1currents}, the hadron mass should depend on the mixing angle $\theta$, which can be thus determined with the inputs of the masses of $K_1(1270)$ and $K_1(1400)$. We shall first calculate the mass of $K_1(1270)$ by using the current $J_{1270}^\mu $, and then obtain the mass of $K_1(1400)$ by substituting ${\theta  \to \theta  + \frac{\pi }{2}}$ for $J_{1400}^\mu $. For the current $J_{1270}^\mu $, the two-point correlation function can be written as
     \begin{eqnarray}
       \begin{gathered}
         \Pi _{1270}^{\mu \nu } = {\rm{i}}\int {{{\rm{d}}^d}x} \;{{\rm{e}}^{{\rm{i}}p \cdot x}}\left\langle \Omega  \right|{\mathsf{T}}\left( {J_{1270}^\mu J_{1270}^{\nu \dag }} \right)\left| \Omega  \right\rangle  \hfill \\
         \;\;\;\;\;\;\;\;\;\, = {\rm{i}}\int {{{\rm{d}}^d}x} \;{{\rm{e}}^{{\rm{i}}p \cdot x}}\left\langle \Omega  \right|{\mathsf{T}}\left( {J_A^\mu J_A^{\nu \dag }} \right)\left| \Omega  \right\rangle {\sin ^2}\theta  \hfill \\
         \;\;\;\;\;\;\;\;\;\;\;\;\ + {\rm{i}}\int {{{\rm{d}}^d}x} \;{{\rm{e}}^{{\rm{i}}p \cdot x}}\left\langle \Omega  \right|{\mathsf{T}}\left( {J_A^\mu J_B^{\nu \dag }} \right)\left| \Omega  \right\rangle \sin \theta \cos \theta  \hfill \\
         \;\;\;\;\;\;\;\;\;\;\;\;\ + {\rm{i}}\int {{{\rm{d}}^d}x} \;{{\rm{e}}^{{\rm{i}}p \cdot x}}\left\langle \Omega  \right|{\mathsf{T}}\left( {J_B^\mu J_A^{\nu \dag }} \right)\left| \Omega  \right\rangle \sin \theta \cos \theta  \hfill \\
         \;\;\;\;\;\;\;\;\;\;\;\;\ + {\rm{i}}\int {{{\rm{d}}^d}x} \;{{\rm{e}}^{{\rm{i}}p \cdot x}}\left\langle \Omega  \right|{\mathsf{T}}\left( {J_B^\mu J_B^{\nu \dag }} \right)\left| \Omega  \right\rangle {\cos ^2}\theta  \hfill \\
         \;\;\;\;\;\;\;\;\;\, \equiv \Pi _{AA}^{\mu \nu }{\sin ^2}\theta  + \Pi _{AB}^{\mu \nu }\sin \theta \cos \theta  \hfill \\
         \;\;\;\;\;\;\;\;\;\;\;\;\ + \Pi _{BA}^{\mu \nu }\sin \theta \cos \theta  + \Pi _{BB}^{\mu \nu }{\cos ^2}\theta .\; \hfill \\ 
       \end{gathered}  \label{correlationfunction}
     \end{eqnarray}

In Ref.~\cite{Dag:2012zz}, the authors defined the current $J_B^\mu (x)\equiv J_B^{\mu \nu }{p_\nu } = \lambda_B\bar s{\sigma ^{\mu \nu }}{{\rm i}p_\nu }\gamma_5 q$, where $p_\nu$ is the external momentum. However, they didn't provide the mass sum rule analyses by using such currents. Instead, they calculated the mixing angle by assuming the orthogonality of the mass eigenstates~\cite{Dag:2012zz}. Actually, the two-point QCD sum rule analyses show that it is difficult to reproduce the masses of $K_1(1270)$ and $K_1(1400)$ mesons no matter what value of the mixing angle adopted. 

In this work, we introduce the following construction of $J_B^\mu (x)$
\begin{eqnarray}
J_B^\mu  \equiv {R_B}J_B^{\mu \nu }{x_\nu } = {R_B}{\bar s}{\sigma ^{\mu \nu }}{x_\nu }\gamma_5q,\;
\end{eqnarray}
where the constant ${R_B}$ is used to compensate the dimension in the mixed currents.

\subsection{Formalism of QCD Sum Rules \label{subB}} 
QCD sum rule has been proven to be a very powerful non-perturbative method for studying hadron properties~\cite{Shifman:1978bx,Reinders:1984sr,Colangelo:2000dp}. 
Using the mixed interpolating currents in Eq.~\eqref{K1currents}, we can calculate the two-point correlation functions for the axial-vector $K_1$ mesons as
\smallskip
     \begin{eqnarray}
       \begin{gathered}
\Pi_X^{\mu \nu }\left( {{p^2}} \right) = {\rm{i}}\int {{{\rm{d}}^d}x} \;{{\rm{e}}^{{\rm{i}}p \cdot x}}\left\langle \Omega  \right|{\mathsf{T}}\left[ {J_X^\mu (x)J_X^{\nu \dag }(0)} \right]\left| \Omega  \right\rangle  \hfill \\
\;\;\;\;\;\;\;\;\;\;\;\;\;\;= \frac{{{p_\mu }{p_\nu }}}{{{p^2}}}\Pi_X^{\rm{S}}\left( {{p^2}} \right) + \left( {\frac{{{p_\mu }{p_\nu }}}{{{p^2}}} - {g_{\mu \nu }}} \right)\Pi_X^{\rm{V}}\left( {{p^2}} \right),\; \hfill \\ 
       \end{gathered} 
     \end{eqnarray}
where $\Pi_X^{\rm{S}}(p^2)$ and $\Pi_X^{\rm{V}}(p^2)$ correspond to the spin-0 and spin-1 intermediate states, respectively. In this work, we shall study the $\Pi_X^{\rm{V}}(p^2)$ invariant structure to investigate the $K_1$ mesons. The mixed currents in Eq.~\eqref{K1currents} can couple to the axial-vector states via
\begin{align}\label{eq: couplingrelation1}
        \left\langle \Omega \left\lvert J_{\mu}\right\rvert X\right\rangle =f_X\epsilon_{\mu}\,,
\end{align}
where $\epsilon_{\mu}$ is the polarization vector and $f_X$ is the coupling constant.

At the hadronic level, the correlation function can be expressed in the form of the dispersion relation
\smallskip
     \begin{eqnarray}
       \begin{gathered}
         \Pi_X^{\rm{V}}\left( {{p^2}} \right) = \frac{1}{\pi }\int_{{s_{\min }}}^\infty  {{\rm{d}}s} \;\frac{{\operatorname{Im} \Pi_X^{\rm{V}}\left( s \right)}}{{s - {p^2} - {\rm{i}}\varepsilon }},\;
       \end{gathered} 
     \end{eqnarray}
     where $s_{\min}=(m_s+m_q)^2$ denotes the physical threshold. 
The imaginary part of the correlation function is usually defined as the spectral function as 
\begin{equation}
\begin{aligned}
    \rho(s)\equiv\frac{1}{\pi}\text{Im} \Pi(s)&=\sum_{n}\delta(s-m_n^2)\langle \Omega\lvert J\rvert n\rangle \langle n \lvert J^{\dagger} \rvert \Omega \rangle \\ &=f_X^2\delta(s-m_X^2)+\cdots,
\end{aligned}
\end{equation}
where the narrow resonance approximation is adopted in the last step, and ``$\cdots$'' contains contributions from the continuum and higher excited states.

We shall  perform the Borel transform on the correlation function to suppress the continuum and higher excited states contributions 
     \begin{eqnarray}
      \Pi _{X}^{\rm{V}}\left( {{M^2_{\rm{B}}}} \right) \equiv \int_{{s_{\min }}}^{\infty} {{\rm{d}}s} \;{{\rm{e}}^{ -s/M_{\rm B}^2}\rho(s)},\;
     \end{eqnarray}
in which $M_{\rm B}$ is the Borel parameter. 
     
We use the method of operator product expansion (OPE) to calculate the two-point correlation function and spectral function at the quark-gluon level. To calculate the Wilson coefficients, we adopt the $d$-Dimensional coordinate space expression for the light quark full propagator  
     \begin{eqnarray}
       \begin{gathered} \label{propagator}
         S_{ab}^q(x) \equiv \langle \Omega |{\mathsf{T}}\left[ {{q_a}(x){{{\bar q}}_b}(0)} \right]\left| \Omega  \right\rangle  \hfill \\
         \;\;\;\;\;\;\;\;\;\;\; = \frac{{{\rm{i}}{\delta _{ab}}\Gamma \left( {\frac{d}{2}} \right)\slashed{x}}}{{2{\pi ^{\frac{d}{2}}}{{\left( { - {x^2}} \right)}^{\frac{d}{2}}}}} + \frac{{{m_q}{\delta _{ab}}\Gamma \left( {\frac{d}{2} - 1} \right)}}{{4{\pi ^{\frac{d}{2}}}{{\left( { - {x^2}} \right)}^{\frac{d}{2} - 1}}}} - \frac{{{\delta _{ab}}}}{{12}}\left\langle {\bar qq} \right\rangle  \hfill \\
         \;\;\;\;\;\;\;\;\;\;\;\;\;\;\;\; + \frac{{{\rm{i}}{m_q}{\delta _{ab}}\slashed{x}}}{{48}}\left\langle {\bar qq} \right\rangle  + \frac{{{\delta _{ab}}{x^2}}}{{192}}\left\langle {g{\bar q}\sigma  \cdot {Gq}} \right\rangle  \hfill \\
         \;\;\;\;\;\;\;\;\;\;\;\;\;\;\;\; - \frac{{{\rm{i}}{m_q}{\delta _{ab}}{x^2}\slashed{x}}}{{{2^7} \cdot {3^2}}}\left\langle {g{\bar q}\sigma  \cdot {Gq}} \right\rangle  - \frac{{{\rm{i}}{\delta _{ab}}{x^2}\slashed{x}}}{{{2^5} \cdot {3^5}}}{g^2}{\left\langle {\bar qq} \right\rangle ^2} \hfill \\
         \;\;\;\;\;\;\;\;\;\;\;\;\;\;\;\; - \frac{{{\rm{i}}\Gamma \left( {\frac{d}{2} - 1} \right)\left( {{\sigma _{\mu \nu }}\slashed{x} + \slashed{x}{\sigma _{\mu \nu }}} \right)}}{{32{\pi ^{\frac{d}{2}}}{{\left( { - {x^2}} \right)}^{\frac{d}{2} - 1}}}}gG_\alpha ^{\mu \nu }\frac{{\lambda _{ab}^\alpha }}{2},\; \hfill \\ 
       \end{gathered} 
     \end{eqnarray}
in which $\slashed{x}=\gamma_\mu x^\mu$, $\Gamma $ is the Gamma function, and the subscripts $a, b$ are color indices. The first two terms in Eq.~\eqref{propagator} are the free quark propagator, while the rest terms represent various nonperturbative contributions. 

\begin{figure}[H]
  \centering 
  \setlength{\abovecaptionskip}{1pt}
  \setlength{\belowcaptionskip}{-10pt}
  \includegraphics[width=0.48\textwidth]{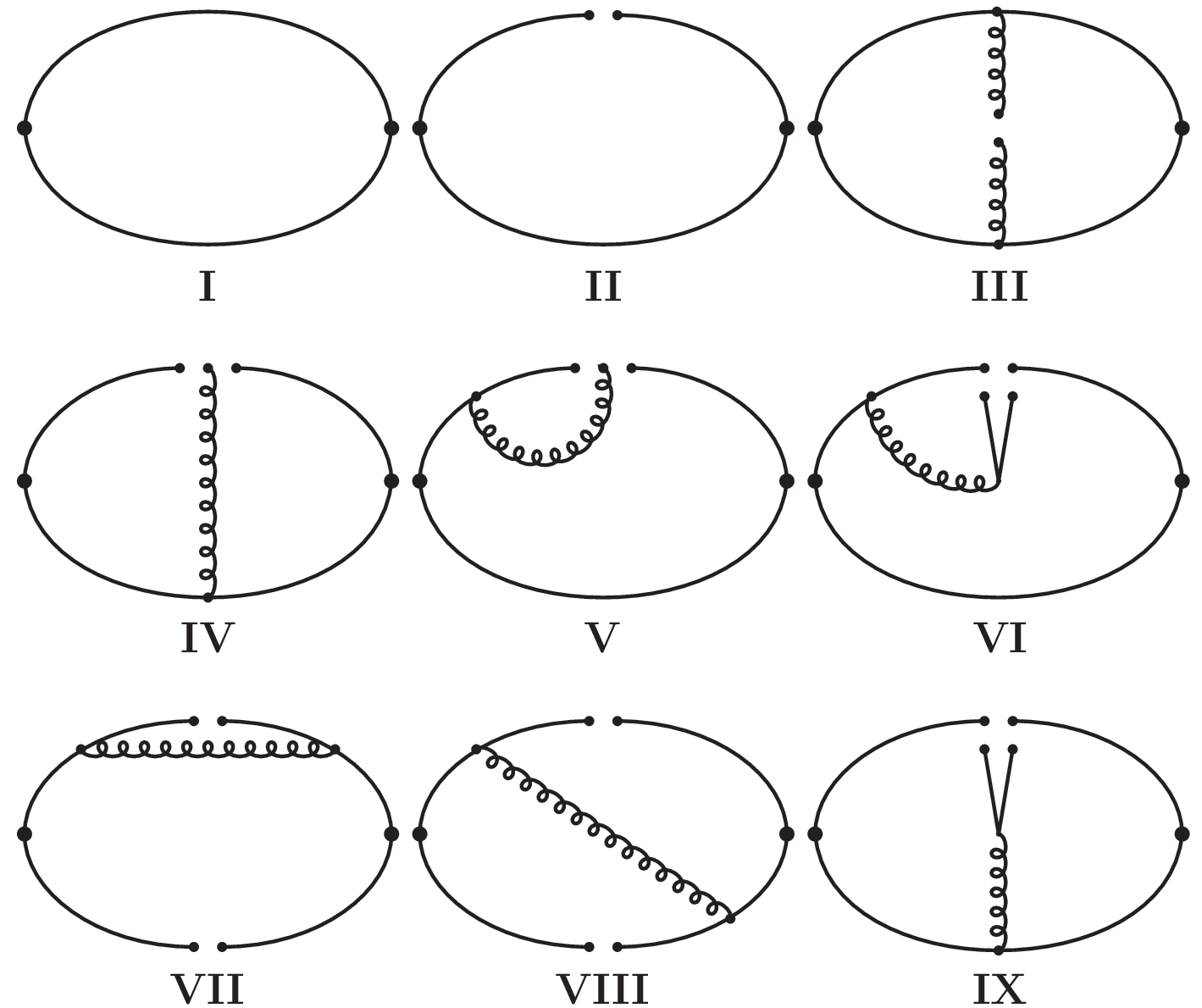}
  \caption{\label{K_1-Feynman-diagrams} Feynman diagrams considered in the calculation of the two-point correlation functions.}
\end{figure}

For the two-point correlation function, we consider the nonperturbative terms including the quark condensates, gluon condensates, quark-gluon mixed condensates and four-quark condensates, as shown in Fig.~\ref{K_1-Feynman-diagrams}. The results show that the diagrams IV and V will cancel out with each other, so that there is no contribution from the mixed condensates at the leading order of $\alpha_{\rm s}$. We calculate the four-quark condensates corresponding to the Feynman diagrams VI, VII, VIII, and IX in Fig.~\ref{K_1-Feynman-diagrams}. Such four-quark condensates have been proven to be very important in the $q\bar q$ light meson sum rules~\cite{Shifman:1978bx,Reinders:1984sr}. The calculations of these diagrams are quite complicated. For convenience, we show the one-to-one correspondence expressions in the Appendix.
For the current $J_{1270}^\mu (x)$, we show the expression of the two-point correlation function as 
\bigskip
\begin{widetext}
  \begin{flalign}\label{pi1270}
\begin{gathered}
  \Pi _{1270}^{\rm{V}}({p^2}) =  - \frac{{{p^2}{{\sin }^2}\theta }}{{4{\pi ^2}}}\ln \left( { - {p^2}} \right) + \frac{{3{m_q}{m_s}{{\sin }^2}\theta }}{{4{\pi ^2}}}\ln \left( { - {p^2}} \right) + \frac{{9R_B^2{{\cos }^2}\theta }}{{4{\pi ^2}}}\ln \left( { - {p^2}} \right) \hfill \\
  \;\;\;\;\;\;\;\;\;\;\;\;\;\;\;\;\;\;\;\, + \frac{{{{\sin }^2}\theta }}{{48{\pi ^2}{p^2}}}\left\langle {{g^2}{G^2}} \right\rangle  - \frac{{{m_s}{{\sin }^2}\theta }}{{{p^2}}}\left\langle {\bar qq} \right\rangle  - \frac{{{m_q}{{\sin }^2}\theta }}{{{p^2}}}\left\langle {\bar ss} \right\rangle  - \frac{{3{m_q}{m_s}R_B^2{{\cos }^2}\theta }}{{2{\pi ^2}{p^2}}} \hfill \\
  \;\;\;\;\;\;\;\;\;\;\;\;\;\;\;\;\;\;\;\, - \frac{{R_B^2{{\cos }^2}\theta }}{{48{\pi ^2}{p^4}}}\left\langle {{g^2}{G^2}} \right\rangle  - \frac{{8{g^2}{{\sin }^2}\theta }}{{81{p^4}}}{\left\langle {\bar qq} \right\rangle ^2} - \frac{{8{g^2}{{\sin }^2}\theta }}{{9{p^4}}}\left\langle {\bar qq} \right\rangle \left\langle {\bar ss} \right\rangle  + \frac{{{m_q}R_B^2{{\cos }^2}\theta }}{{{p^4}}}\left\langle {\bar qq} \right\rangle  \hfill \\
  \;\;\;\;\;\;\;\;\;\;\;\;\;\;\;\;\;\;\;\, + \frac{{2{m_s}R_B^2{{\cos }^2}\theta }}{{{p^4}}}\left\langle {\bar qq} \right\rangle  - \frac{{8{g^2}{{\sin }^2}\theta }}{{81{p^4}}}{\left\langle {\bar ss} \right\rangle ^2} + \frac{{2{m_q}R_B^2{{\cos }^2}\theta }}{{{p^4}}}\left\langle {\bar ss} \right\rangle  + \frac{{{m_s}R_B^2{{\cos }^2}\theta }}{{{p^4}}}\left\langle {\bar ss} \right\rangle  \hfill \\
  \;\;\;\;\;\;\;\;\;\;\;\;\;\;\;\;\;\;\;\, + \frac{{8{g^2}{m_s}{R_B}\sin \theta \cos \theta }}{{9{p^6}}}{\left\langle {\bar qq} \right\rangle ^2} - \frac{{4{g^2}{m_q}{m_s}{{\sin }^2}\theta }}{{9{p^6}}}\left\langle {\bar qq} \right\rangle \left\langle {\bar ss} \right\rangle  \hfill \\
  \;\;\;\;\;\;\;\;\;\;\;\;\;\;\;\;\;\;\;\, + \frac{{16{g^2}R_B^2{{\cos }^2}\theta }}{{3{p^6}}}\left\langle {\bar qq} \right\rangle \left\langle {\bar ss} \right\rangle  + \frac{{8{g^2}{m_q}{R_B}\sin \theta \cos \theta }}{{9{p^6}}}{\left\langle {\bar ss} \right\rangle ^2}.\; \hfill \\ 
\end{gathered}
  \end{flalign} 
\end{widetext}
\bigskip
According to Eq.~\eqref{K1currents}, the correlation function for $J_{1400}^\mu (x)$ can be obtained directly by taking replacement $\theta\to\theta+\pi/2$ in Eq.~\eqref{pi1270}. 

Via the quark-hadron duality, the QCD sum rules can be established by combining the correlation functions at both hadronic and quark-gluonic levels
\begin{equation}
f_X^2m_X^2{{\rm{e}}^{ - m_X^2/M_{\rm{B}}^2}} = \int_{{s_{\min }}}^{{s_0}} {{\rm{d}}s} \;{{\rm{e}}^{ - s/M_{\rm{B}}^2}}\rho (s){\mkern 1mu} \,,
\end{equation}
where $s_0$ is the continuum threshold. The hadron mass of the lowest-lying hadron state can be obtained as 
     \begin{eqnarray} \label{hadronmass}
       {m_X}\left( {M_{\rm{B}}^2,{s_0}} \right) = \sqrt {\frac{{\int_{{s_{\min }}}^{{s_0}} {{\rm{d}}s} \;{{\rm{e}}^{ - s/M_{\rm{B}}^2}}s\rho (s)}}{{\int_{{s_{\min }}}^{{s_0}} {{\rm{d}}s} \;{{\rm{e}}^{ - s/M_{\rm{B}}^2}}\rho (s)}}} .
     \end{eqnarray}

\subsection{Mixing angle of $K_1(1270)$ and $K_1(1400)$}
In this subsection, we determine the mixing angle of the strange axial-vector mesons $K_1(1270)$ and $K_1(1400)$ by reproducing their hadron masses. We shall perform the numerical analyses of the mass sum rules by using the following parameters~\cite{ParticleDataGroup:2022pth, Reinders:1984sr, Narison:2011rn, Kuhn:2007vp}:
     \begin{eqnarray} \label{parameter1}
       \begin{gathered}
         {g^2} = 4\pi \alpha_{\rm s}  = 5.29_{ - 0.16}^{ + 0.17},\; \hfill \\
         {m_q} = 3.45_{ - 0.15}^{ + 0.35} \times {10^{ - 3}}\;{\rm{GeV}},\; \hfill \\
         {m_s} = 93.4_{ - 3.4}^{ + 8.6} \times {10^{ - 3}}\;{\rm{GeV}},\; \hfill \\
         {m_{1270,{\rm{phy}}}} = \left( {1.253 \pm 0.007} \right)\;{\rm{GeV}},\;\hfill \\
         {m_{1400,{\rm{phy}}}} = \left( {1.403 \pm 0.007} \right)\;{\rm{GeV}},\; \hfill \\
         \left\langle {\bar qq} \right\rangle  =  - {\left( {0.24 \pm 0.01} \right)^3}\;{\rm{GeV}^3},\; \hfill \\
         \left\langle {\bar ss} \right\rangle  = \left( {0.8 \pm 0.1} \right)\left\langle {\bar qq} \right\rangle ,\; \hfill \\
         \left\langle {{g^2}{G^2}} \right\rangle  = \left( {0.48 \pm 0.14} \right)\;{\rm{GeV}^4},\; \hfill \\
         \left\langle {g{\bar q}\sigma  \cdot {Gq}} \right\rangle  =  - \left( {0.8 \pm 0.2} \right)\;{\rm{GeV}^2}\left\langle {\bar qq} \right\rangle ,\; \hfill \\
         \left\langle {g{\bar s}\sigma  \cdot {Gs}} \right\rangle  =  - \left( {0.8 \pm 0.2} \right)\;{\rm{GeV}^2}\left\langle {\bar ss} \right\rangle ,\; \hfill \\ 
       \end{gathered} 
     \end{eqnarray} 
where both ${m_s}$ and ${m_q}$ are the $\overline {{\rm{MS}}} $ masses at the scale $\mu  = 2\;{\rm{GeV}}$. Using the two-loop renormalization group equations, the strong coupling constant is determined by evolution from $\alpha_{\rm s} \left( {{m_Z}} \right)$~\cite{ParticleDataGroup:2022pth}.

\begin{figure}[H]
  \centering 
  \setlength{\abovecaptionskip}{1pt}
  \setlength{\belowcaptionskip}{-10pt}
  \includegraphics[width=0.48\textwidth]{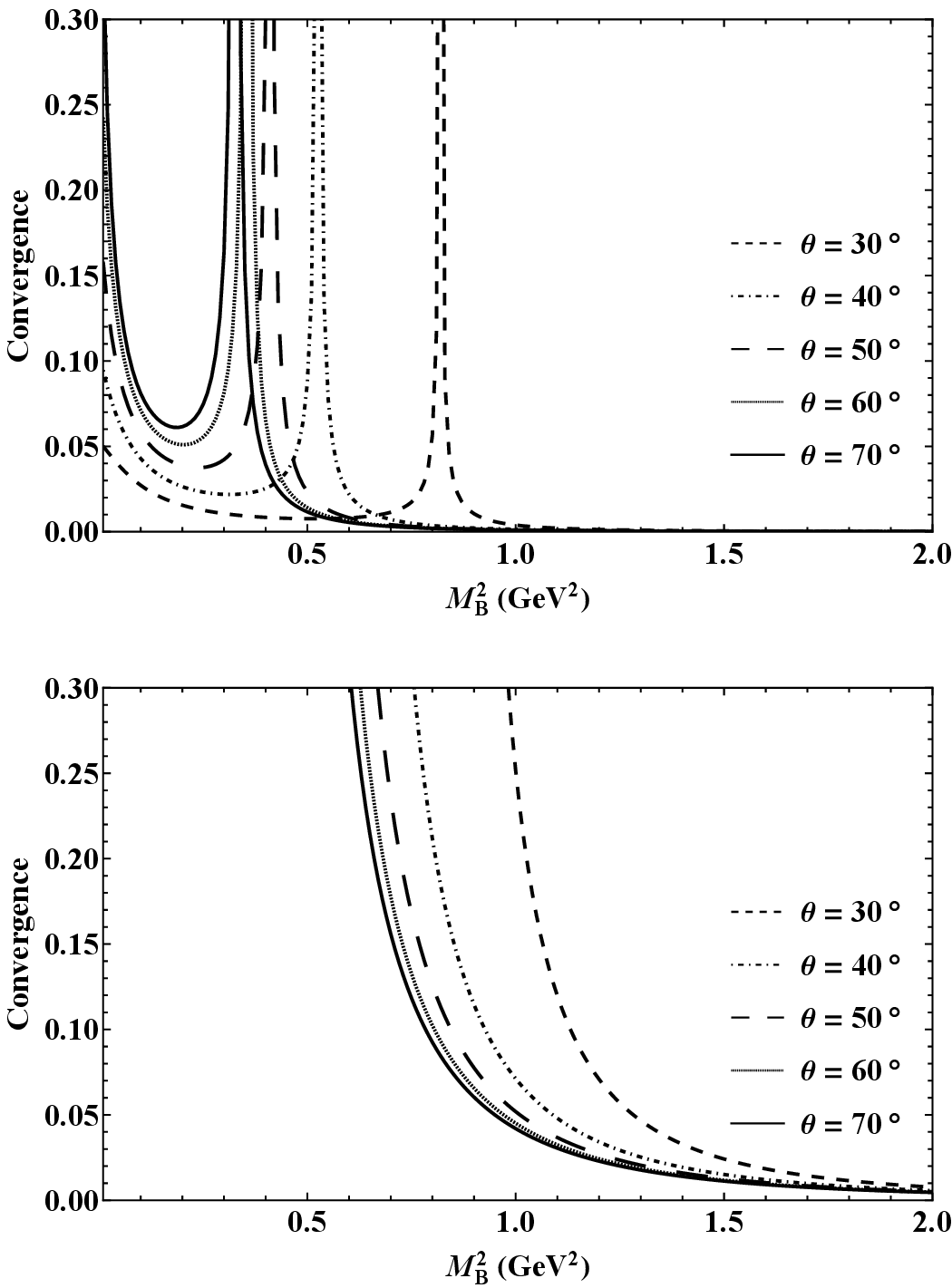}
  \caption{\label{K_1-Converg-compare} OPE convergences without (top) and with (bottom) considering the contributions from four-quark condensates for the current $J_{1270}^\mu (x)$.}
\end{figure}

As shown in Eq.~\eqref{hadronmass}, the hadron mass is the function of the Borel mass $M_{\rm B}$ and continuum threshold $s_0$. We study the OPE convergence (CVG) and pole contribution function (PC) to determine the lower and upper bounds of $M_{\rm B}^2$ respectively
     \begin{eqnarray} \label{CVG}
       {\rm{CVG}}\left( {{M^2_{\rm{B}}}} \right) &\equiv & \left| {\frac{{\Pi^{{\rm{V,}}\dim 6}\left( {{M^2_{\rm{B}}},\infty } \right)}}{{\Pi^{\rm{V}}\left( {{M^2_{\rm{B}}},\infty } \right)}}} \right|,\; \\
       {\rm{PC}}\left( {{M^2_{\rm{B}}},{s_0}} \right) &\equiv & \frac{{\Pi^{\rm{V}}\left( {{M^2_{\rm{B}}},{s_0}} \right)}}{{\Pi^{\rm{V}}\left( {{M^2_{\rm{B}}},\infty } \right)}},\; \label{PC}
     \end{eqnarray} 
where $\Pi^{{\rm{V,}}\dim 6}$ represents the contributions of dimension-6  four-quark condensates. We choose the optimal value of the continuum threshold $s_0$ 
to minimize the variance of the ground state hadron mass with respect to ${M_{\rm{B}}}$. For the current $J_{1270}^\mu (x)$, 
we show the CVG behaviors without and with the contributions from four-quark condensates in Fig.~\ref{K_1-Converg-compare}. 
It is clear that these four-quark condensates can improve the OPE convergence, as mentioned in Sec.~\ref{subB}. 

The parameters ${{R_B}}$ and $\theta $ will be fine-tuned to reproduce the masses of $K_1(1270)$ and $K_1(1400)$. For the fixed value of ${R_B}$, 
one only needs to study the variation of ${m_{1270}}$ with respect to the mixing angle $\theta $, since the mass of $K_1(1400)$ can be determined by ${m_{1400}}\left( \theta  \right) = {m_{1270}}\left( {\theta  + \frac{\pi }{2}} \right)$.

\begin{figure}[H]
  \centering 
  \setlength{\abovecaptionskip}{1pt}
  \setlength{\belowcaptionskip}{-10pt}
  \includegraphics[width=0.48\textwidth]{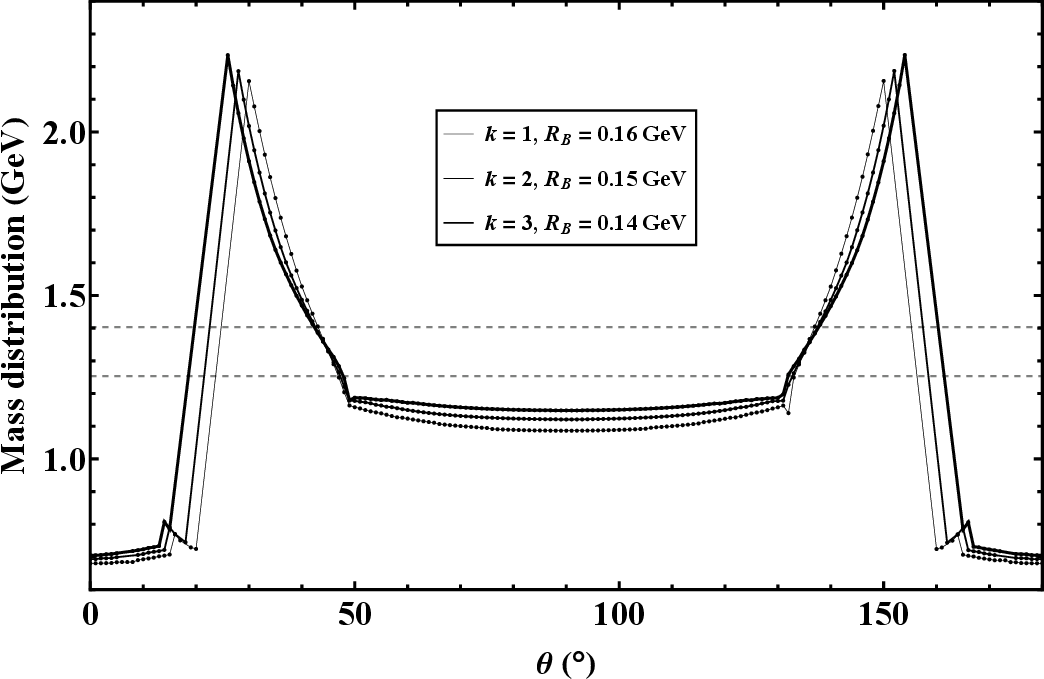}
  \caption{\label{K_1-Mass-distribution} Dependences of the mass distributions on the mixing angle.} 
\end{figure}

In Fig.~\ref{K_1-Mass-distribution}, we study the mass distribution depending on the mixing angle for ${R_B}=0.16$ GeV, by requiring that PC $\geq 50\% $ and CVG $\leq10\% $. It is shown that the masses of $K_1(1270)$ and $K_1(1400)$ can be reproduced in the distribution for $\theta  = {46.95^ \circ }$ and $\theta= {136.95^ \circ }$ respectively. Considering the violation of the factorization assumption with a factor $k\sim 1-3$, the lower bound of the Borel window will be slightly raised in our analyses. We also show the mass distributions for $k=2, 3$ in Fig.~\ref{K_1-Mass-distribution}, from which one finds that the influence for the mixing angle is quite small. Therefore, we choose $k=1$ in our following analyses.

Considering both currents in Eq.~\eqref{K1currents}, the parameters ${{R_B}}$ and $\theta $ are finally obtained as 
     \begin{eqnarray}\label{RBtheta}
      {R_B} = (0.16 \pm 0.01)~\text{GeV},\;\theta  = \left( {46.95_{ - 0.23}^{ + 0.25}} \right)^\circ ,\;
     \end{eqnarray} 
where the errors in ${R_B}$ primarily originate from $\left\langle {{g^2}{G^2}} \right\rangle $ 
     and $\left\langle {\bar qq} \right\rangle $, while the errors in $\theta $ mainly arise from $\left\langle {\bar qq} \right\rangle $, $\left\langle {\bar ss} \right\rangle $, $\left\langle {{g^2}{G^2}} \right\rangle $, ${m_{1270,{\rm{phy}}}}$ and ${m_{1400,{\rm{phy}}}}$ in Eq.~\eqref{parameter1}. 
Our prediction of the mixing angle is consistent with the results of Refs.~\cite{Blundell:1995au,Burakovsky:1997dd}.

\begin{figure}[H]
  \centering 
  \setlength{\abovecaptionskip}{1pt}
  \setlength{\belowcaptionskip}{-10pt}
  \includegraphics[width=0.48\textwidth]{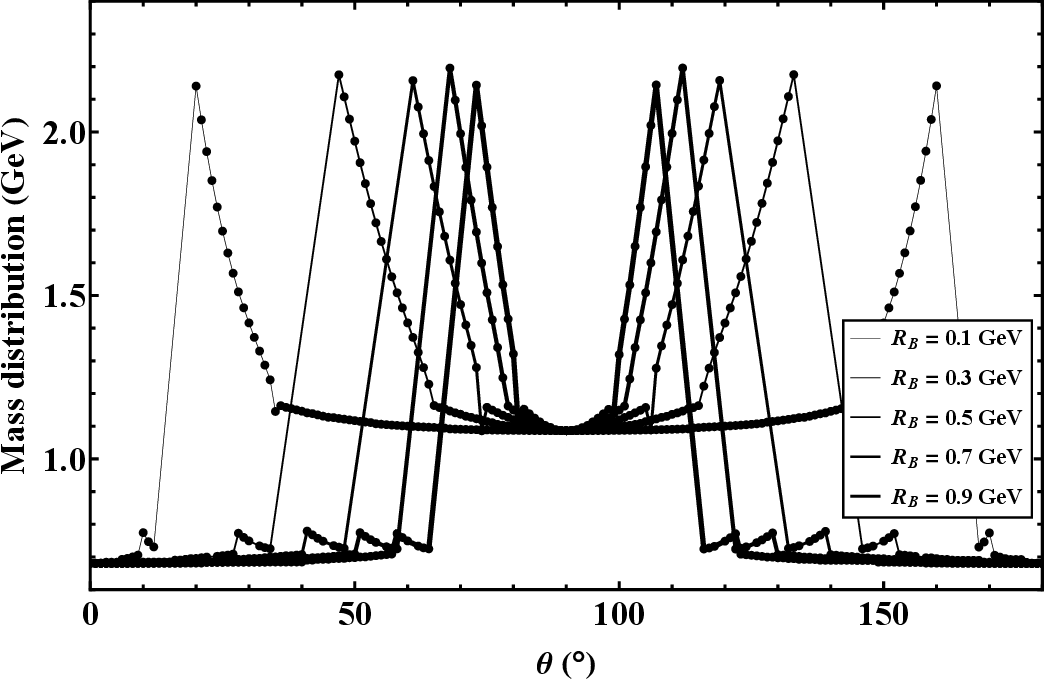}
  \caption{\label{K_1-Different-RB-values} Mass distributions for different values of ${R_B}$.} 
\end{figure}

In Fig.~\ref{K_1-Different-RB-values}, we show the mass distributions with different values of the parameter ${R_B}$, which display very similar behaviors with the same requirements PC $\geq 50\% $ and CVG $\leq10\% $. To maintain the mass relation ${m_{1400}}\left( \theta  \right) = {m_{1270}}\left( {\theta  + \frac{\pi }{2}} \right)$, the values of ${R_B}$ and $\theta $ should be uniquely determined as in Eq.~\eqref{RBtheta}.

We also study the dependences of ${R_B}$ and $\theta $ on different requirements of CVG and PC in Eqs.~\eqref{CVG}-\eqref{PC}. As shown in Table~\ref{tab:K_1-Tab_1}, the values of ${R_B}$ and $\theta $ only change slightly with variations of CVG and PC, providing reasonable parameter ranges for the mass distributions. 
\begin{table}[H]
\caption{\label{tab:K_1-Tab_1} Dependences of ${R_B}$ and $\theta $ on different requirements of CVG and PC.}
\begin{ruledtabular}
\begin{tabular}{cccc}
\textrm{CVG}&
\textrm{PC}&
\textrm{${R_B}$}&
\textrm{$\theta $}\\
\colrule
$1\%  $ & $20\% $ & 0.19 GeV & ${46.83^ \circ }$\\
$5\%  $ & $40\% $ & 0.17 GeV & ${46.90^ \circ }$\\
$10\% $ & $50\% $ & 0.16 GeV & ${46.95^ \circ }$\\
$15\% $ & $60\% $ & 0.15 GeV & ${47.03^ \circ }$\\
$20\% $ & $70\% $ & 0.15 GeV & ${47.14^ \circ }$\\
\end{tabular}
\end{ruledtabular}
\end{table}

\section{$K\bar K_1(1270/1400)$ molecular states }
To investigate the $K\bar K_1(1400)$ molecular interpretation of $\eta_1(1855)$, 
we try to construct the interpolating current for the $K\bar K_1(1400)$ molecular state with ${J^{PC}} = {1^{ -+ }}$. Using the operator of $K_1(1270)$ in Eq.~\eqref{K1currents}, we firstly construct the current for $K\bar K_1(1270)$ state 
     \begin{eqnarray}\label{KK_1}
       \begin{gathered}
         J_{K\bar K_1(1270)}^\mu (x) = {{{\bar s}}_a}(x){\Gamma _K}{q_a}(x){{{\bar q}}_b}(x)\bar \Gamma _A^\mu {{s}_b}(x)\sin \theta  \hfill \\
         \;\;\;\;\;\;\;\;\;\;\;\;\;\;\;\;\;\;\;\;\;\;\;\;\;\, + {{{\bar s}}_a}(x){\Gamma _K}{q_a}(x){{{\bar q}}_b}(x)\bar \Gamma _B^\mu {{s}_b}(x)\cos \theta  \hfill \\
         \;\;\;\;\;\;\;\;\;\;\;\;\;\;\;\;\;\;\;\;\;\;\;\;\;\, + {{{\bar q}}_a}(x){{\bar \Gamma }_K}{{s}_a}(x){{{\bar s}}_b}(x)\Gamma _A^\mu {q_b}(x)\sin \theta  \hfill \\
         \;\;\;\;\;\;\;\;\;\;\;\;\;\;\;\;\;\;\;\;\;\;\;\;\;\, + {{{\bar q}}_a}(x){{\bar \Gamma }_K}{{s}_a}(x){{{\bar s}}_b}(x)\Gamma _B^\mu {q_b}(x)\cos \theta ,\; \hfill \\ 
       \end{gathered} 
     \end{eqnarray}
where 
\begin{eqnarray}
\nonumber {\Gamma _K} &=& {\rm{i}}{\gamma_5},\;{{\bar \Gamma }_K} = {\gamma ^0}\Gamma _K^\dag {\gamma ^0} = {\rm{i}}{\gamma_5}, \Gamma _A^\mu  = {\gamma ^\mu }{\gamma _5},\\ 
\bar \Gamma _A^\mu  &=& {\gamma ^0}\Gamma _A^{\mu \dag }{\gamma ^0} =  - {\gamma _5}{\gamma ^\mu },  \Gamma _B^\mu  = {R_B}{\sigma ^{\mu \nu }}{x_\nu }{\gamma _5},\\ \nonumber 
\bar \Gamma _B^\mu  &=& {\gamma ^0}\Gamma _B^{\mu \dag }{\gamma ^0} =  - {R_B}{\gamma _5}{\sigma ^{\mu \nu }}{x_\nu }, 
\end{eqnarray} 
in which the parameter $R_B$ and mixing angle $\theta$ are determined in Eq.~\eqref{RBtheta}. 
The interpolating current $J_{K\bar K_1\left( {1400} \right)}^\mu $ for the $K\bar K_1(1400)$ molecular state can be obtained by taking the replacement $\theta \to\theta  +\pi/2$ in $J_{K\bar K_1\left( {1270} \right)}^\mu $. We calculate the two-point correlation functions for these two molecular currents up to dimension-9 condensates at the leading order of $\alpha_{\rm s}$.
The detailed expression for the correlation function $\Pi _{K\bar K_1\left( {1270} \right)}^{\rm{V}}$ is given as
\begin{widetext} 
  \begin{flalign}
    \Pi _{K\bar K_1\left( {1270} \right)}^{\rm{V}} = \ln \left( { - {p^2}} \right)\Pi _{K\bar K_1\left( {1270} \right)}^{{\rm{V,}}\;{\rm{ln}}\left( { - {p^2}} \right)} + \frac{1}{{{p^2}}}\Pi _{K\bar K_1\left( {1270} \right)}^{{\rm{V,}}\;{p^{ - 2}}} + \frac{1}{{{p^4}}}\Pi _{K\bar K_1\left( {1270} \right)}^{{\rm{V,}}\;{p^{ - 4}}},\;
  \end{flalign}
  \par\noindent where 
  \begin{flalign}
    &\begin{gathered}
       \Pi _{K{{\bar K}_1}\left( {1270} \right)}^{{\text{V}},\;{\text{ln}}\left( { - {p^2}} \right)} \equiv  - \frac{{{p^8}{{\sin }^2}\theta }}{{49152{\pi ^6}}} + \frac{{9R_B^2{p^6}{{\cos }^2}\theta }}{{5120{\pi ^6}}} - \frac{{{m_q}{p^4}{{\sin }^2}\theta }}{{128{\pi ^4}}}\left\langle {\bar qq} \right\rangle  + \frac{{5{m_q}R_B^2{p^2}{{\cos }^2}\theta }}{{32{\pi ^4}}}\left\langle {\bar qq} \right\rangle  - \frac{{{m_s}{p^4}{{\sin }^2}\theta }}{{256{\pi ^4}}}\left\langle {\bar qq} \right\rangle  \hfill \\
       \;\;\;\;\;\;\;\;\;\;\;\;\;\;\;\;\;\;\;\;\;\; - \frac{{{m_q}{p^4}{{\sin }^2}\theta }}{{256{\pi ^4}}}\left\langle {\bar ss} \right\rangle  - \frac{{{m_s}{p^4}{{\sin }^2}\theta }}{{128{\pi ^4}}}\left\langle {\bar ss} \right\rangle  + \frac{{5{m_s}R_B^2{p^2}{{\cos }^2}\theta }}{{32{\pi ^4}}}\left\langle {\bar ss} \right\rangle  - \frac{{{p^4}{{\sin }^2}\theta }}{{12288{\pi ^6}}}\left\langle {{g^2}{G^2}} \right\rangle  \hfill \\
       \;\;\;\;\;\;\;\;\;\;\;\;\;\;\;\;\;\;\;\;\;\; + \frac{{5R_B^2{p^2}{{\cos }^2}\theta }}{{1536{\pi ^6}}}\left\langle {{g^2}{G^2}} \right\rangle  - \frac{{{m_q}{p^2}{{\sin }^2}\theta }}{{384{\pi ^4}}}\left\langle {g\bar q\sigma  \cdot Gq} \right\rangle  - \frac{{{m_s}{p^2}{{\sin }^2}\theta }}{{256{\pi ^4}}}\left\langle {g\bar q\sigma  \cdot Gq} \right\rangle  \hfill \\
       \;\;\;\;\;\;\;\;\;\;\;\;\;\;\;\;\;\;\;\;\;\; - \frac{{{m_q}{p^2}{{\sin }^2}\theta }}{{256{\pi ^4}}}\left\langle {g\bar s\sigma  \cdot Gs} \right\rangle  - \frac{{{m_s}{p^2}{{\sin }^2}\theta }}{{384{\pi ^4}}}\left\langle {g\bar s\sigma  \cdot Gs} \right\rangle  + \frac{{{p^2}{{\sin }^2}\theta }}{{96{\pi ^2}}}{\left\langle {\bar qq} \right\rangle ^2} + \frac{{{p^2}{{\sin }^2}\theta }}{{24{\pi ^2}}}\left\langle {\bar qq} \right\rangle \left\langle {\bar ss} \right\rangle  \hfill \\
       \;\;\;\;\;\;\;\;\;\;\;\;\;\;\;\;\;\;\;\;\;\; + \frac{{{p^2}{{\sin }^2}\theta }}{{96{\pi ^2}}}{\left\langle {\bar ss} \right\rangle ^2} + \frac{{{m_q}{{\sin }^2}\theta }}{{1024{\pi ^4}}}\left\langle {{g^2}{G^2}} \right\rangle \left\langle {\bar qq} \right\rangle  - \frac{{5{m_s}{{\sin }^2}\theta }}{{1024{\pi ^4}}}\left\langle {{g^2}{G^2}} \right\rangle \left\langle {\bar qq} \right\rangle  - \frac{{5{m_q}{{\sin }^2}\theta }}{{1024{\pi ^4}}}\left\langle {{g^2}{G^2}} \right\rangle \left\langle {\bar ss} \right\rangle  \hfill \\
       \;\;\;\;\;\;\;\;\;\;\;\;\;\;\;\;\;\;\;\;\;\; + \frac{{{m_s}{{\sin }^2}\theta }}{{1024{\pi ^4}}}\left\langle {{g^2}{G^2}} \right\rangle \left\langle {\bar ss} \right\rangle  + \frac{{{{\sin }^2}\theta }}{{64{\pi ^2}}}\left\langle {g\bar q\sigma  \cdot Gq} \right\rangle \left\langle {\bar ss} \right\rangle  + \frac{{{{\sin }^2}\theta }}{{64{\pi ^2}}}\left\langle {\bar qq} \right\rangle \left\langle {g\bar s\sigma  \cdot Gs} \right\rangle  + \frac{{{{\sin }^2}\theta }}{{16384{\pi ^6}}}{\left\langle {{g^2}{G^2}} \right\rangle ^2}, \hfill \\ 
     \end{gathered}&
  \end{flalign}
 
  \begin{flalign}
    &\begin{gathered}
       \Pi _{K{{\bar K}_1}\left( {1270} \right)}^{{\text{V}},\;{p^{ - 2}}} \equiv \frac{{{m_q}R_B^2{{\cos }^2}\theta }}{{384{\pi ^4}}}\left\langle {{g^2}{G^2}} \right\rangle \left\langle {\bar qq} \right\rangle  + \frac{{{m_s}R_B^2{{\cos }^2}\theta }}{{96{\pi ^4}}}\left\langle {{g^2}{G^2}} \right\rangle \left\langle {\bar qq} \right\rangle  + \frac{{{m_q}R_B^2{{\cos }^2}\theta }}{{96{\pi ^4}}}\left\langle {{g^2}{G^2}} \right\rangle \left\langle {\bar ss} \right\rangle  \hfill \\
       \;\;\;\;\;\;\;\;\;\;\;\;\;\;\;\;\;\;\;\;\;\; + \frac{{{m_s}R_B^2{{\cos }^2}\theta }}{{384{\pi ^4}}}\left\langle {{g^2}{G^2}} \right\rangle \left\langle {\bar ss} \right\rangle  - \frac{{R_B^2{{\cos }^2}\theta }}{{12288{\pi ^6}}}{\left\langle {{g^2}{G^2}} \right\rangle ^2} - \frac{{11{m_q}{{\sin }^2}\theta }}{{9216{\pi ^4}}}\left\langle {{g^2}{G^2}} \right\rangle \left\langle {g\bar q\sigma  \cdot Gq} \right\rangle  \hfill \\
       \;\;\;\;\;\;\;\;\;\;\;\;\;\;\;\;\;\;\;\;\;\; - \frac{{5{m_s}{{\sin }^2}\theta }}{{3072{\pi ^4}}}\left\langle {{g^2}{G^2}} \right\rangle \left\langle {g\bar q\sigma  \cdot Gq} \right\rangle  - \frac{{5{m_q}{{\sin }^2}\theta }}{{3072{\pi ^4}}}\left\langle {{g^2}{G^2}} \right\rangle \left\langle {g\bar s\sigma  \cdot Gs} \right\rangle  - \frac{{11{m_s}{{\sin }^2}\theta }}{{9216{\pi ^4}}}\left\langle {{g^2}{G^2}} \right\rangle \left\langle {g\bar s\sigma  \cdot Gs} \right\rangle  \hfill \\
       \;\;\;\;\;\;\;\;\;\;\;\;\;\;\;\;\;\;\;\;\;\; + \frac{{{m_q}{{\sin }^2}\theta }}{6}{\left\langle {\bar qq} \right\rangle ^2}\left\langle {\bar ss} \right\rangle  - \frac{{{m_s}{{\sin }^2}\theta }}{4}{\left\langle {\bar qq} \right\rangle ^2}\left\langle {\bar ss} \right\rangle  - \frac{{{m_q}{{\sin }^2}\theta }}{4}\left\langle {\bar qq} \right\rangle {\left\langle {\bar ss} \right\rangle ^2} + \frac{{{m_s}{{\sin }^2}\theta }}{6}\left\langle {\bar qq} \right\rangle {\left\langle {\bar ss} \right\rangle ^2}, \hfill \\ 
     \end{gathered} &
  \end{flalign}

  \begin{flalign}
    &\begin{gathered}
      \Pi _{K{{\bar K}_1}\left( {1270} \right)}^{{\text{V}},\;{p^{ - 4}}} \equiv \frac{{17{m_q}R_B^2{{\cos }^2}\theta }}{{9216{\pi ^4}}}\left\langle {{g^2}{G^2}} \right\rangle \left\langle {g\bar q\sigma  \cdot Gq} \right\rangle  + \frac{{{m_s}R_B^2{{\cos }^2}\theta }}{{384{\pi ^4}}}\left\langle {{g^2}{G^2}} \right\rangle \left\langle {g\bar q\sigma  \cdot Gq} \right\rangle  + \frac{{{m_q}R_B^2{{\cos }^2}\theta }}{{384{\pi ^4}}}\left\langle {{g^2}{G^2}} \right\rangle \left\langle {g\bar s\sigma  \cdot Gs} \right\rangle  \hfill \\
      \;\;\;\;\;\;\;\;\;\;\;\;\;\;\;\;\;\;\;\;\;\; + \frac{{17{m_s}R_B^2{{\cos }^2}\theta }}{{9216{\pi ^4}}}\left\langle {{g^2}{G^2}} \right\rangle \left\langle {g\bar s\sigma  \cdot Gs} \right\rangle  + \frac{{2{m_s}R_B^2{{\cos }^2}\theta }}{3}{\left\langle {\bar qq} \right\rangle ^2}\left\langle {\bar ss} \right\rangle  + \frac{{2{m_q}R_B^2{{\cos }^2}\theta }}{3}\left\langle {\bar qq} \right\rangle {\left\langle {\bar ss} \right\rangle ^2}. \hfill \\ 
    \end{gathered}&
  \end{flalign}
\end{widetext}

To perform QCD sum rule analyses for the $K\bar K_1$ molecular systems,  we define similar CVG and PC functions as in Eqs.~\eqref{CVG}-\eqref{PC} to determine reasonable working regions of Borel mass and continuum threshold 
     \begin{equation}
       {\rm{CVG}_{K\bar K_1}}\left( {{M_{\rm{B}}}} \right) \equiv \left| {\frac{{\Pi _{K\bar K_1}^{{\rm{V}},\dim 8}\left( {{M_{\rm{B}}},\infty } \right)}}{{\Pi _{K\bar K_1}^{\rm{V}}\left( {{M_{\rm{B}}},\infty } \right)}}} \right|,\;
     \end{equation}
     \begin{equation}
       {\rm{P}}{{\rm{C}}_{K\bar K_1}}\left( {{M_{\rm B}},{s_0}} \right) \equiv \frac{{\Pi _{K\bar K_1}^{\rm{V}}\left( {{M_{\rm{B}}},{s_0}} \right)}}{{\Pi _{K\bar K_1}^{\rm{V}}\left( {{M_{\rm{B}}},\infty } \right)}}.\;
     \end{equation}

\begin{figure}[H]
  \centering 
  \setlength{\abovecaptionskip}{1pt}
  \setlength{\belowcaptionskip}{-10pt}
  \includegraphics[width=0.48\textwidth]{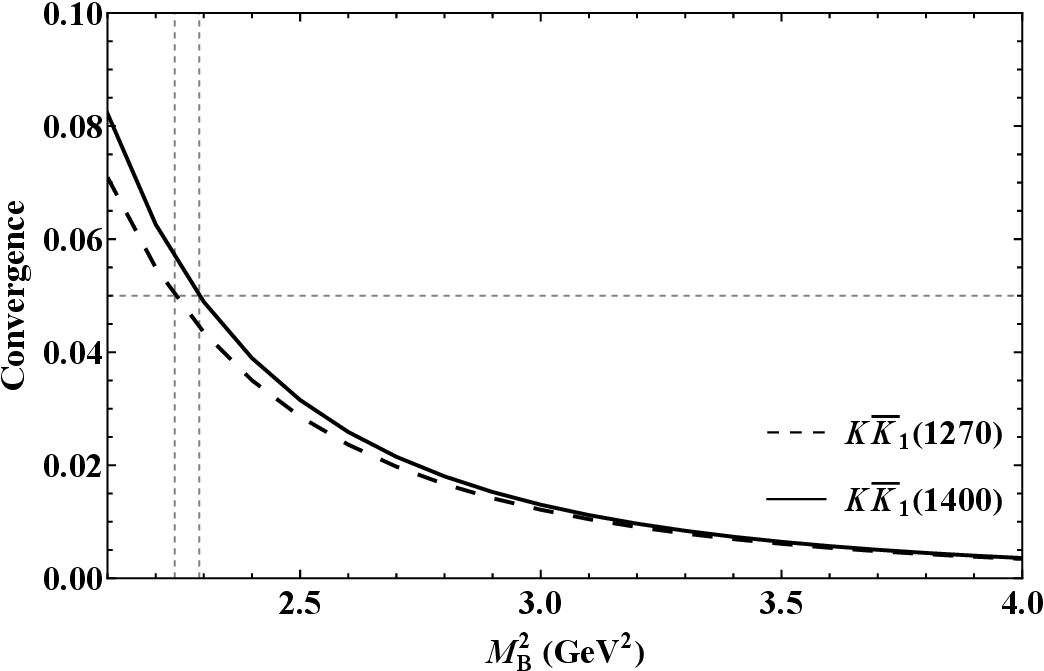}
  \caption{\label{eta_1-ConverPlot} OPE convergences of $K\bar K_1(1270/1400)$ systems.} 
\end{figure}

\begin{figure}[H]
  \centering 
  \setlength{\abovecaptionskip}{1pt}
  \setlength{\belowcaptionskip}{-10pt}
  \includegraphics[width=0.48\textwidth]{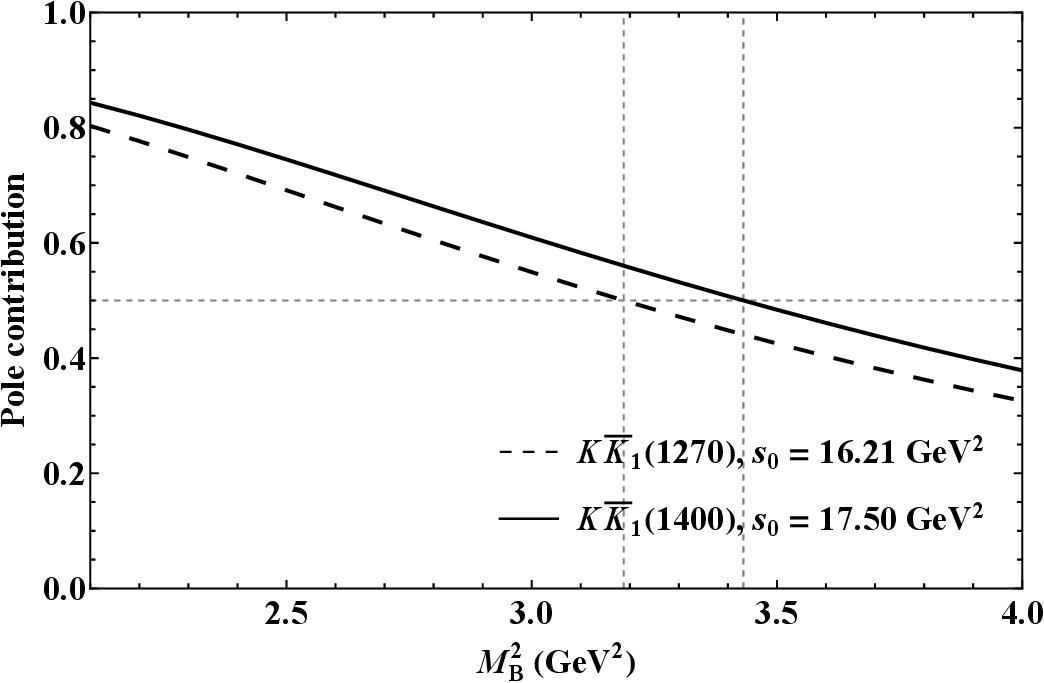}
  \caption{\label{eta_1-PoleConPlot} Pole contributions of $K\bar K_1(1270/1400)$ systems.} 
\end{figure}

\begin{figure}[H]
  \centering 
  \setlength{\abovecaptionskip}{1pt}
  \setlength{\belowcaptionskip}{-10pt}
  \includegraphics[width=0.48\textwidth]{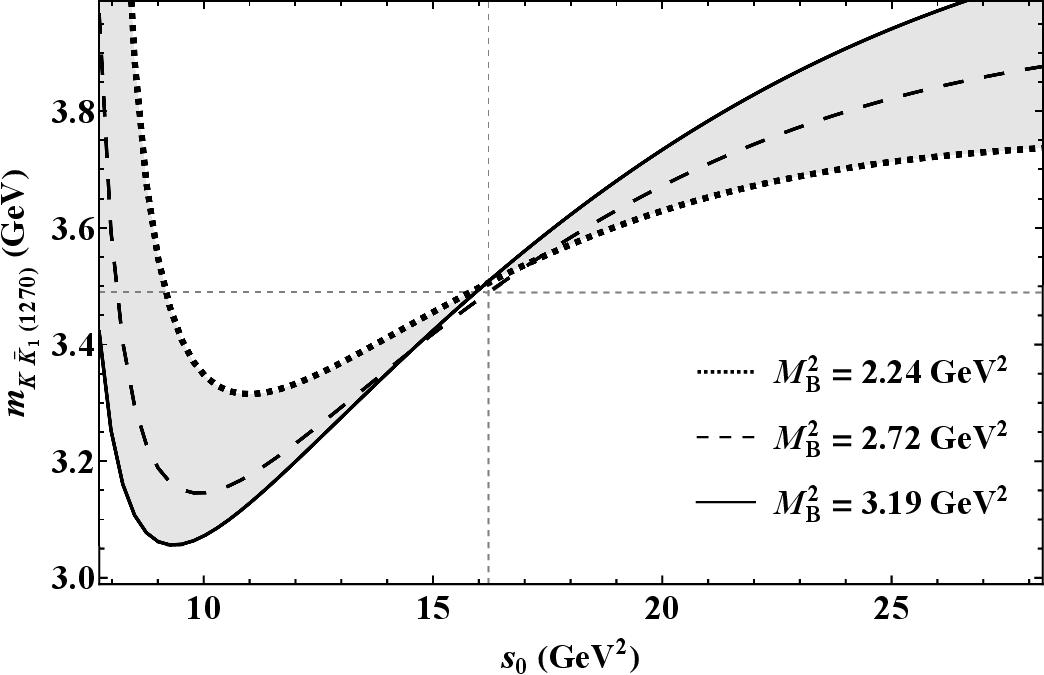}
  \caption{\label{eta_1-Mass-s0-1270} Mass curves for $K\bar K_1(1270)$ state.} 
\end{figure}

\begin{figure}[H]
  \centering 
  \setlength{\abovecaptionskip}{1pt}
  \setlength{\belowcaptionskip}{-10pt}
  \includegraphics[width=0.48\textwidth]{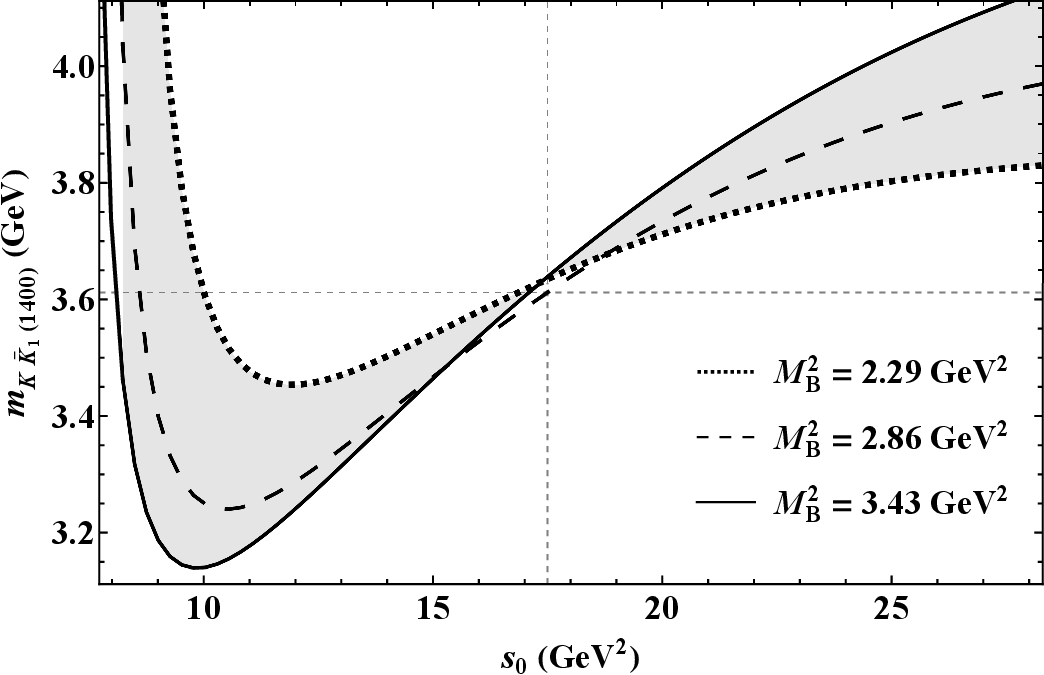}
  \caption{\label{eta_1-Mass-s0-1400} Mass curves for $K\bar K_1(1400)$ state.} 
\end{figure}

In Fig.~\ref{eta_1-ConverPlot} and Fig.~\ref{eta_1-PoleConPlot}, we show the OPE convergences and pole contributions of two molecular systems, from which one can obtain the parameter working regions in mass sum rules. Then we plot the mass curves with respect to the continuum threshold $s_0$ in Fig.~\ref{eta_1-Mass-s0-1270} and Fig.~\ref{eta_1-Mass-s0-1400} for $K\bar K_1(1270)$ and $K\bar K_1(1400)$ molecular states respectively. One finds that the masses for $K\bar K_1(1270/1400)$ molecular states are much higher than the mass of $\eta_1(1855)$. 

\begin{figure}[H]
  \centering 
  \setlength{\textfloatsep}{20pt}
  \setlength{\abovecaptionskip}{1pt}
  \setlength{\belowcaptionskip}{-10pt}
  \includegraphics[width=0.48\textwidth]{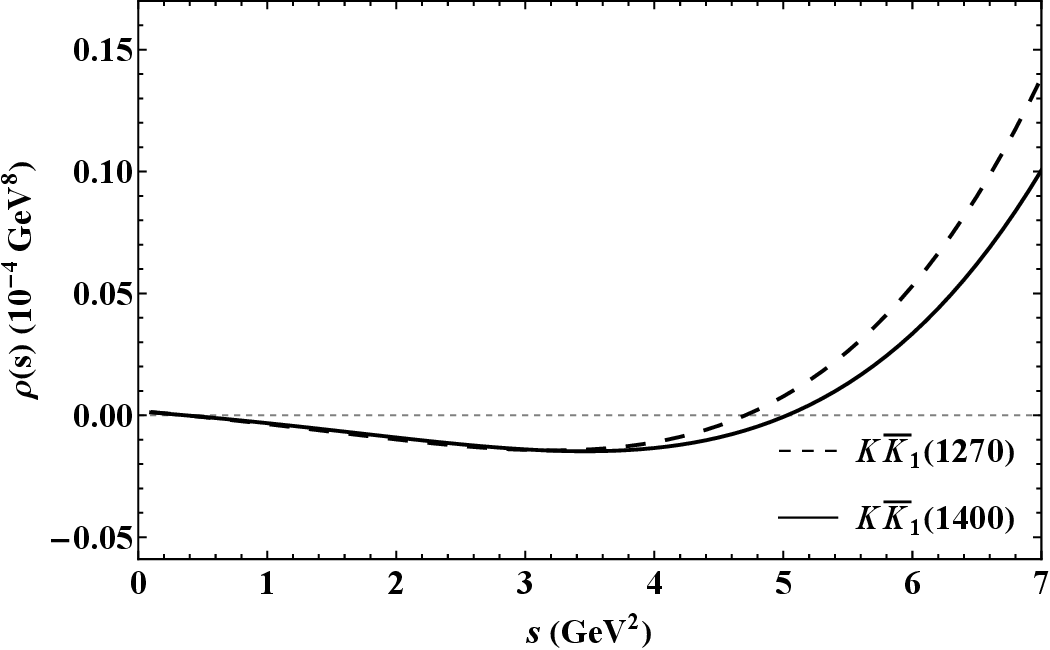}
  \caption{Spectral functions of $K\bar K1(1270/1400)$ systems.} \label{rho(s)_KK_1}
\end{figure}

However, the above mass sum rules analyses based on the interpolating currents $J_{K\bar K_1\left( {1270} \right)}^\mu $ and $J_{K\bar K_1\left( {1400} \right)}^\mu $ are not reasonable due to the bad behaviors of their spectral functions. As shown in Fig.~\ref{rho(s)_KK_1}, both the spectral functions become negative in the regions $s<5$ GeV$^2$, implying that they are not able to provide reliable investigations of the $K\bar K_1$ molecular states. 
\vspace{3pt}

\section{Summary}
\par In this work, we propose a new construction for the operators of axial-vector $K_1(1270)$ and $K_1(1400)$ mesons, which are the mixtures of P-wave $K_{1A}\left( {^3{P_1}} \right)$ and $K_{1B}\left( {^1{P_1}} \right)$ states. 
We calculate their two-point correlation functions by including the dimension-6 four-quark condensates at the next-to-leading order of $\alpha_{\rm s}$, which are proven to be very important for the OPE convergence and mass sum rules stabilities. We determine the mixing angle $\theta  = \left( {46.95_{ - 0.23}^{ + 0.25}} \right)^\circ$ by reproducing the masses of $K_1(1270)$ and $K_1(1400)$ mesons, which is consistent with the results of Refs.~\cite{Blundell:1995au,Burakovsky:1997dd}.

To explore the molecular interpretation of $\eta_1(1855)$, we construct the corresponding molecular interpolating currents for $K\bar K_1(1270)$ and $K\bar K_1(1400)$ states with $J^{PC}=1^{-+}$. We calculate the correlation functions and investigate their mass sum rules. However, the spectral functions from these two currents $J_{K\bar K_1\left( {1270} \right)}^\mu $ and $J_{K\bar K_1\left( {1400} \right)}^\mu $ show bad behavior of the positivity so that they can not give reliable mass predictions. 

\section*{Acknowledgments}
This work is supported by the National Natural Science Foundation of China under Grant No. 12175318, the Natural Science Foundation of Guangdong Province of China under Grants No. 2022A1515011922 and No. 2023A1515011704.

\bibliographystyle{paper}
\bibliography{References}

\appendix
\begin{widetext}
\section*{Appendix: Correlation functions of the four-quark condensates for the current $J_{1270}^\mu (x)$}\label{appendix}
For the current $J_{1270}^\mu (x)$, we show the correlation functions corresponding to Feynman diagrams VII, VIII, and IX in Fig.~\ref{K_1-Feynman-diagrams} as the following
  \begin{flalign}
    &\begin{gathered}
      \Pi _{{\rm{VII}} - 1}^{\mu \nu }({p^2}) =  - {g^2}\frac{{\left\langle {\bar qq} \right\rangle \left\langle {\bar ss} \right\rangle }}{{36}}\frac{1}{{{p^2} - m_{s}^2 + {\rm{i}}\varepsilon }}\frac{{\rm{1}}}{{{p^2} - m_{s}^2 + {\rm{i}}\varepsilon }}\frac{1}{{{p^2} + {\rm{i}}\varepsilon }}{\rm{tr}}\left[ {\Gamma _A^\mu \bar \Gamma _A^\nu \left( { - \slashed{p} + {m_s}} \right){\gamma ^\eta }{\gamma _\eta }\left( { - \slashed{p} + {m_s}} \right)} \right]{\sin ^2}\theta  \hfill \\
      \;\;\;\;\;\;\;\;\;\;\;\;\;\;\;\;\;\;\;\;\;\; -{\rm{i}}{g^2}\frac{{\left\langle {\bar qq} \right\rangle \left\langle {\bar ss} \right\rangle }}{{36}}\int {{{\rm{d}}^4}k} \;\delta \left( {k - p} \right)\frac{\partial }{{\partial {k^\sigma }}}\left\{ \begin{gathered}
      \;\;\;\frac{1}{{{k^2} - m_{s}^2 + {\rm{i}}\varepsilon }}\frac{{\rm{1}}}{{{k^2} - m_{s}^2 + {\rm{i}}\varepsilon }}\frac{1}{{{k^2} + {\rm{i}}\varepsilon }} \hfill \\
      \times {\rm{tr}}\left[ {\Gamma _A^\mu \frac{{\bar \Gamma _B^\nu }}{{{x_\sigma }}}\left( { - \slashed{k} + {m_s}} \right){\gamma ^\eta }{\gamma _\eta }\left( { - \slashed{k} + {m_s}} \right)} \right] \hfill \\ 
      \end{gathered}  \right\}\sin \theta \cos \theta  \hfill \\
      \;\;\;\;\;\;\;\;\;\;\;\;\;\;\;\;\;\;\;\;\;\; -{\rm{i}}{g^2}\frac{{\left\langle {\bar qq} \right\rangle \left\langle {\bar ss} \right\rangle }}{{36}}\int {{{\rm{d}}^4}k} \;\delta \left( {k - p} \right)\frac{\partial }{{\partial {k^\rho }}}\left\{ \begin{gathered}
      \;\;\;\frac{1}{{{k^2} - m_{s}^2 + {\rm{i}}\varepsilon }}\frac{{\rm{1}}}{{{k^2} - m_{s}^2 + {\rm{i}}\varepsilon }}\frac{1}{{{k^2} + {\rm{i}}\varepsilon }} \hfill \\
      \times {\rm{tr}}\left[ {\frac{{\Gamma _B^\mu }}{{{x_\rho }}}\bar \Gamma _A^\nu \left( { - \slashed{k} + {m_s}} \right){\gamma ^\eta }{\gamma _\eta }\left( { - \slashed{k} + {m_s}} \right)} \right] \hfill \\ 
      \end{gathered}  \right\}\cos \theta \sin \theta  \hfill \\
      \;\;\;\;\;\;\;\;\;\;\;\;\;\;\;\;\;\;\;\;\;\; -g^2\frac{{\left\langle {\bar qq} \right\rangle \left\langle {\bar ss} \right\rangle }}{{36}}\int {{{\rm{d}}^4}k} \;\delta \left( {k - p} \right)\frac{\partial }{{\partial {k^\rho }}}\frac{\partial }{{\partial {k^\sigma }}}\left\{ \begin{gathered}
      \;\;\;\frac{1}{{{k^2} - m_{s}^2 + {\rm{i}}\varepsilon }}\frac{{\rm{1}}}{{{k^2} - m_{s}^2 + {\rm{i}}\varepsilon }}\frac{1}{{{k^2} + {\rm{i}}\varepsilon }} \hfill \\
      \times {\rm{tr}}\left[ {\frac{{\Gamma _B^\mu }}{{{x_\rho }}}\frac{{\bar \Gamma _B^\nu }}{{{x_\sigma }}}\left( { - \slashed{k} + {m_s}} \right){\gamma ^\eta }{\gamma _\eta }\left( { - \slashed{k} + {m_s}} \right)} \right] \hfill \\ 
      \end{gathered}  \right\}{\cos ^2}\theta ,\; \hfill \\ 
    \end{gathered}&
  \end{flalign}

  \begin{flalign}
    &\begin{gathered}
      \Pi _{{\rm{VII}} - 2}^{\mu \nu }({p^2}) =  - {g^2}\frac{{\left\langle {\bar ss} \right\rangle \left\langle {\bar qq} \right\rangle }}{{36}}\frac{{\rm{1}}}{{{p^2} + {\rm{i}}\varepsilon }}\frac{{\rm{1}}}{{{p^2} - m_q^2 + {\rm{i}}\varepsilon }}\frac{{\rm{1}}}{{{p^2} - m_q^2 + {\rm{i}}\varepsilon }}{\rm{tr}}\left[ {\Gamma _A^\mu \left( {\slashed{p} + {m_q}} \right){\gamma _\eta }{\gamma ^\eta }\left( {\slashed{p} + {m_q}} \right)\bar \Gamma _A^\nu } \right]{\sin ^2}\theta  \hfill \\
      \;\;\;\;\;\;\;\;\;\;\;\;\;\;\;\;\;\;\;\;\;\; -{\rm{i}}{g^2}\frac{{\left\langle {\bar ss} \right\rangle \left\langle {\bar qq} \right\rangle }}{{36}}\int {{{\rm{d}}^4}k} \;\delta \left( {k - p} \right)\frac{\partial }{{\partial {k^\sigma }}}\left\{ \begin{gathered}
      \;\;\;\frac{{\rm{1}}}{{{k^2} + {\rm{i}}\varepsilon }}\frac{{\rm{1}}}{{{k^2} - m_q^2 + {\rm{i}}\varepsilon }}\frac{{\rm{1}}}{{{k^2} - m_q^2 + {\rm{i}}\varepsilon }} \hfill \\
      \times {\rm{tr}}\left[ {\Gamma _A^\mu \left( {\slashed{k} + {m_q}} \right){\gamma _\eta }{\gamma ^\eta }\left( {\slashed{k} + {m_q}} \right)\frac{{\bar \Gamma _B^\nu }}{{{x_\sigma }}}} \right] \hfill \\ 
      \end{gathered}  \right\}\sin \theta \cos \theta  \hfill \\
      \;\;\;\;\;\;\;\;\;\;\;\;\;\;\;\;\;\;\;\;\;\; -{\rm{i}}{g^2}\frac{{\left\langle {\bar ss} \right\rangle \left\langle {\bar qq} \right\rangle }}{{36}}\int {{{\rm{d}}^4}k} \;\delta \left( {k - p} \right)\frac{\partial }{{\partial {k^\rho }}}\left\{ \begin{gathered}
      \;\;\;\frac{{\rm{1}}}{{{k^2} + {\rm{i}}\varepsilon }}\frac{{\rm{1}}}{{{k^2} - m_q^2 + {\rm{i}}\varepsilon }}\frac{{\rm{1}}}{{{k^2} - m_q^2 + {\rm{i}}\varepsilon }} \hfill \\
      \times {\rm{tr}}\left[ {\frac{{\Gamma _B^\mu }}{{{x_\rho }}}\left( {\slashed{k} + {m_q}} \right){\gamma _\eta }{\gamma ^\eta }\left( {\slashed{k} + {m_q}} \right)\bar \Gamma _A^\nu } \right] \hfill \\ 
      \end{gathered}  \right\}\cos \theta \sin \theta  \hfill \\
      \;\;\;\;\;\;\;\;\;\;\;\;\;\;\;\;\;\;\;\;\;\; -g^2\frac{{\left\langle {\bar ss} \right\rangle \left\langle {\bar qq} \right\rangle }}{{36}}\int {{{\rm{d}}^4}k} \;\delta \left( {k - p} \right)\frac{\partial }{{\partial {k^\rho }}}\frac{\partial }{{\partial {k^\sigma }}}\left\{ \begin{gathered}
      \;\;\;\frac{{\rm{1}}}{{{k^2} + {\rm{i}}\varepsilon }}\frac{{\rm{1}}}{{{k^2} - m_q^2 + {\rm{i}}\varepsilon }}\frac{{\rm{1}}}{{{k^2} - m_q^2 + {\rm{i}}\varepsilon }} \hfill \\
      \times {\rm{tr}}\left[ {\frac{{\Gamma _B^\mu }}{{{x_\rho }}}\left( {\slashed{k} + {m_q}} \right){\gamma _\eta }{\gamma ^\eta }\left( {\slashed{k} + {m_q}} \right)\frac{{\bar \Gamma _B^\nu }}{{{x_\sigma }}}} \right] \hfill \\ 
      \end{gathered}  \right\}{\cos ^2}\theta ,\; \hfill \\ 
    \end{gathered}&
  \end{flalign}

  \begin{flalign}
    &\begin{gathered}
       \Pi _{{\rm{VIII}} - 1}^{\mu \nu }({p^2}) =  - {g^2}\frac{{\left\langle {\bar ss} \right\rangle \left\langle {{\rm{qq}}} \right\rangle }}{{36}}\frac{{\rm{1}}}{{{p^2} + {\rm{i}}\varepsilon }}\frac{{\rm{1}}}{{{p^2} - m_{s}^2 + {\rm{i}}\varepsilon }}\frac{{\rm{1}}}{{{p^2} - m_q^2 + {\rm{i}}\varepsilon }}{\rm{tr}}\left[ {\Gamma _A^\mu {\gamma ^\eta }\left( {\slashed{p} + {m_q}} \right)\bar \Gamma _A^\nu {\gamma _\eta }\left( { - \slashed{p} + {m_s}} \right)} \right]{\sin ^2}\theta  \hfill \\
       \;\;\;\;\;\;\;\;\;\;\;\;\;\;\;\;\;\;\;\;\;\;\; -{\rm{i}}{g^2}\frac{{\left\langle {\bar ss} \right\rangle \left\langle {{\rm{qq}}} \right\rangle }}{{36}}\int {{{\rm{d}}^4}k} \;\delta \left( {k - p} \right)\frac{\partial }{{\partial {k^\sigma }}}\left\{ \begin{gathered}
       \;\;\;\frac{{\rm{1}}}{{{k^2} + {\rm{i}}\varepsilon }}\frac{{\rm{1}}}{{{k^2} - m_{s}^2 + {\rm{i}}\varepsilon }}\frac{{\rm{1}}}{{{k^2} - m_q^2 + {\rm{i}}\varepsilon }} \hfill \\
       \times {\rm{tr}}\left[ {\Gamma _A^\mu {\gamma ^\eta }\left( {\slashed{k} + {m_q}} \right)\frac{{\bar \Gamma _B^\nu }}{{{x_\sigma }}}{\gamma _\eta }\left( { - \slashed{k} + {m_s}} \right)} \right] \hfill \\ 
       \end{gathered}  \right\}\sin \theta \cos \theta  \hfill \\
       \;\;\;\;\;\;\;\;\;\;\;\;\;\;\;\;\;\;\;\;\;\;\; -{\rm{i}}{g^2}\frac{{\left\langle {\bar ss} \right\rangle \left\langle {{\rm{qq}}} \right\rangle }}{{36}}\int {{{\rm{d}}^4}k} \;\delta \left( {k - p} \right)\frac{\partial }{{\partial {k^\rho }}}\left\{ \begin{gathered}
       \;\;\;\frac{{\rm{1}}}{{{k^2} + {\rm{i}}\varepsilon }}\frac{{\rm{1}}}{{{k^2} - m_{s}^2 + {\rm{i}}\varepsilon }}\frac{{\rm{1}}}{{{k^2} - m_q^2 + {\rm{i}}\varepsilon }} \hfill \\
       \times {\rm{tr}}\left[ {\frac{{\Gamma _B^\mu }}{{{x_\rho }}}{\gamma ^\eta }\left( {\slashed{k} + {m_q}} \right)\bar \Gamma _A^\nu {\gamma _\eta }\left( { - \slashed{k} + {m_s}} \right)} \right] \hfill \\ 
       \end{gathered}  \right\}\cos \theta \sin \theta  \hfill \\
       \;\;\;\;\;\;\;\;\;\;\;\;\;\;\;\;\;\;\;\;\;\;\; -g^2\frac{{\left\langle {\bar ss} \right\rangle \left\langle {{\rm{qq}}} \right\rangle }}{{36}}\int {{{\rm{d}}^4}k} \;\delta \left( {k - p} \right)\frac{\partial }{{\partial {k^\rho }}}\frac{\partial }{{\partial {k^\sigma }}}\left\{ \begin{gathered}
       \;\;\;\frac{{\rm{1}}}{{{k^2} + {\rm{i}}\varepsilon }}\frac{{\rm{1}}}{{{k^2} - m_{s}^2 + {\rm{i}}\varepsilon }}\frac{{\rm{1}}}{{{k^2} - m_q^2 + {\rm{i}}\varepsilon }} \hfill \\
       \times {\rm{tr}}\left[ {\frac{{\Gamma _B^\mu }}{{{x_\rho }}}{\gamma ^\eta }\left( {\slashed{k} + {m_q}} \right)\frac{{\bar \Gamma _B^\nu }}{{{x_\sigma }}}{\gamma _\eta }\left( { - \slashed{k} + {m_s}} \right)} \right] \hfill \\ 
       \end{gathered}  \right\}{\cos ^2}\theta ,\; \hfill \\ 
     \end{gathered}&
  \end{flalign}

  \begin{flalign}
    &\begin{gathered}
       \Pi _{{\rm{VIII}} - 2}^{\mu \nu }({p^2}) =  - {g^2}\frac{{\left\langle {\bar qq} \right\rangle \left\langle {\bar ss} \right\rangle }}{{36}}\frac{{\rm{1}}}{{{p^2} + {\rm{i}}\varepsilon }}\frac{1}{{{p^2} - m_q^2 + {\rm{i}}\varepsilon }}\frac{{\rm{1}}}{{{p^2} - m_{s}^2 + {\rm{i}}\varepsilon }}{\rm{tr}}\left[ {{\gamma ^\eta }\Gamma _A^\mu \left( {\slashed{p} + {m_q}} \right){\gamma _\eta }\bar \Gamma _A^\nu \left( { - \slashed{p} + {m_s}} \right)} \right]{\sin ^2}\theta  \hfill \\
       \;\;\;\;\;\;\;\;\;\;\;\;\;\;\;\;\;\;\;\;\;\;\; -{\rm{i}}{g^2}\frac{{\left\langle {\bar qq} \right\rangle \left\langle {\bar ss} \right\rangle }}{{36}}\int {{{\rm{d}}^4}k} \;\delta \left( {k - p} \right)\frac{\partial }{{\partial {k^\sigma }}}\left\{ \begin{gathered}
       \;\;\;\frac{{\rm{1}}}{{{k^2} + {\rm{i}}\varepsilon }}\frac{1}{{{k^2} - m_q^2 + {\rm{i}}\varepsilon }}\frac{{\rm{1}}}{{{k^2} - m_{s}^2 + {\rm{i}}\varepsilon }} \hfill \\
       \times {\rm{tr}}\left[ {{\gamma ^\eta }\Gamma _A^\mu \left( {\slashed{k} + {m_q}} \right){\gamma _\eta }\frac{{\bar \Gamma _B^\nu }}{{{x_\sigma }}}\left( { - \slashed{k} + {m_s}} \right)} \right] \hfill \\ 
       \end{gathered}  \right\}\sin \theta \cos \theta  \hfill \\
       \;\;\;\;\;\;\;\;\;\;\;\;\;\;\;\;\;\;\;\;\;\;\; -{\rm{i}}{g^2}\frac{{\left\langle {\bar qq} \right\rangle \left\langle {\bar ss} \right\rangle }}{{36}}\int {{{\rm{d}}^4}k} \;\delta \left( {k - p} \right)\frac{\partial }{{\partial {k^\rho }}}\left\{ \begin{gathered}
       \;\;\;\frac{{\rm{1}}}{{{k^2} + {\rm{i}}\varepsilon }}\frac{1}{{{k^2} - m_q^2 + {\rm{i}}\varepsilon }}\frac{{\rm{1}}}{{{k^2} - m_{s}^2 + {\rm{i}}\varepsilon }} \hfill \\
       \times {\rm{tr}}\left[ {{\gamma ^\eta }\frac{{\Gamma _B^\mu }}{{{x_\rho }}}\left( {\slashed{k} + {m_q}} \right){\gamma _\eta }\bar \Gamma _A^\nu \left( { - \slashed{k} + {m_s}} \right)} \right] \hfill \\ 
       \end{gathered}  \right\}\cos \theta \sin \theta  \hfill \\
       \;\;\;\;\;\;\;\;\;\;\;\;\;\;\;\;\;\;\;\;\;\;\; -g^2\frac{{\left\langle {\bar qq} \right\rangle \left\langle {\bar ss} \right\rangle }}{{36}}\int {{{\rm{d}}^4}k} \;\delta \left( {k - p} \right)\left( {\frac{\partial }{{\partial {k^\rho }}}\frac{\partial }{{\partial {k^\sigma }}}} \right)\left\{ \begin{gathered}
       \;\;\;\frac{{\rm{1}}}{{{k^2} + {\rm{i}}\varepsilon }}\frac{1}{{{k^2} - m_q^2 + {\rm{i}}\varepsilon }}\frac{{\rm{1}}}{{{k^2} - m_{s}^2 + {\rm{i}}\varepsilon }} \hfill \\
       \times {\rm{tr}}\left[ {{\gamma ^\eta }\frac{{\Gamma _B^\mu }}{{{x_\rho }}}\left( {\slashed{k} + {m_q}} \right){\gamma _\eta }\frac{{\bar \Gamma _B^\nu }}{{{x_\sigma }}}\left( { - \slashed{k} + {m_s}} \right)} \right] \hfill \\ 
       \end{gathered}  \right\}{\cos ^2}\theta ,\; \hfill \\ 
     \end{gathered}&
  \end{flalign}

  \begin{flalign}
     &\begin{gathered}
        \Pi _{{\rm{IX}} - 1}^{\mu \nu }({p^2}) =  - \int {{{\rm{d}}^4}k} \;\delta \left( {k + p} \right){\rm{Pack}}_1\left( {\Gamma _A^\mu ,\bar \Gamma _A^\nu } \right){\sin ^2}\theta  \hfill \\
        \;\;\;\;\;\;\;\;\;\;\;\;\;\;\;\;\;\;\;\;\, - {\rm{i}}\int {{{\rm{d}}^4}k} \;\delta \left( {k + p} \right)\frac{\partial }{{\partial {k^\sigma }}}{\rm{Pack}}_1\left( {\Gamma _A^\mu ,\frac{{\bar \Gamma _B^\nu }}{{{x_\sigma }}}} \right)\sin \theta \cos \theta  \hfill \\
        \;\;\;\;\;\;\;\;\;\;\;\;\;\;\;\;\;\;\;\;\, - {\rm{i}}\int {{{\rm{d}}^4}k} \;\delta \left( {k + p} \right)\frac{\partial }{{\partial {k^\rho }}}{\rm{Pack}}_1\left( {\frac{{\Gamma _B^\mu }}{{{x_\rho }}},\bar \Gamma _A^\nu } \right)\cos \theta \sin \theta  \hfill \\
        \;\;\;\;\;\;\;\;\;\;\;\;\;\;\;\;\;\;\;\;\, + \int {{{\rm{d}}^4}k} \;\delta \left( {k + p} \right)\frac{\partial }{{\partial {k^\rho }}}\frac{\partial }{{\partial {k^\sigma }}}{\rm{Pack}}_1\left( {\frac{{\Gamma _B^\mu }}{{{x_\rho }}},\frac{{\bar \Gamma _B^\nu }}{{{x_\sigma }}}} \right){\cos ^2}\theta ,\; \hfill \\ 
      \end{gathered}&
  \end{flalign}

  \begin{flalign}
     &\begin{gathered}
        \Pi _{{\rm{IX}} - 2}^{\mu \nu }({p^2}) =  - \int {{{\rm{d}}^4}k} \;\delta \left( {k - p} \right){\rm{Pack}_2}\left( {\Gamma _A^\mu ,\bar \Gamma _A^\nu } \right){\sin ^2}\theta  \hfill \\
        \;\;\;\;\;\;\;\;\;\;\;\;\;\;\;\;\;\;\;\;\, + {\rm{i}}\int {{{\rm{d}}^4}k} \;\delta \left( {k - p} \right)\frac{\partial }{{\partial {k^\sigma }}}{\rm{Pack}_2}\left( {\Gamma _A^\mu ,\frac{{\bar \Gamma _B^\nu }}{{{x_\sigma }}}} \right)\sin \theta \cos \theta  \hfill \\
        \;\;\;\;\;\;\;\;\;\;\;\;\;\;\;\;\;\;\;\;\, + {\rm{i}}\int {{{\rm{d}}^4}k} \;\delta \left( {k - p} \right)\frac{\partial }{{\partial {k^\rho }}}{\rm{Pack}_2}\left( {\frac{{\Gamma _B^\mu }}{{{x_\rho }}},\bar \Gamma _A^\nu } \right)\cos \theta \sin \theta  \hfill \\
        \;\;\;\;\;\;\;\;\;\;\;\;\;\;\;\;\;\;\;\;\, + \int {{{\rm{d}}^4}k} \;\delta \left( {k - p} \right)\frac{\partial }{{\partial {k^\rho }}}\frac{\partial }{{\partial {k^\sigma }}}{\rm{Pack}_2}\left( {\frac{{\Gamma _B^\mu }}{{{x_\rho }}},\frac{{\bar \Gamma _B^\nu }}{{{x_\sigma }}}} \right){\cos ^2}\theta ,\; \hfill \\ 
      \end{gathered}&
  \end{flalign}
  and
  \bigskip
  \par\noindent
  $\Gamma _A^\mu  \equiv {\gamma ^\mu }{\gamma _5},\;\bar \Gamma _A^\nu  \equiv  - {\gamma _5}{\gamma ^\nu },\;\frac{{\Gamma _B^\mu }}{{{x_\rho }}} \equiv {R_B}{\sigma ^{\mu \rho }}{\gamma _5},\;\frac{{\bar \Gamma _B^\nu }}{{{x_\sigma }}} \equiv  - {R_B}{\gamma _5}{\sigma ^{\nu \sigma }},\;$
  \bigskip
  \par\noindent 
    $\begin{gathered}
     {\rm{Pack}}_1\left( {{\Gamma _1},{\Gamma _2}} \right) \equiv \int {{{\rm{d}}^4}q} \;\delta \left( {q - k} \right)\frac{\partial }{{\partial {q_\eta }}}\frac{\partial }{{\partial {q_\delta }}}\left\{ \begin{gathered}
     \;\;\;\frac{{{g^2}{{\left\langle {\bar qq} \right\rangle }^2}}}{{324\left( {{k^2} - m_{s}^2 + {\rm{i}}\varepsilon } \right)\left( {{q^2} - m_{s}^2 + {\rm{i}}\varepsilon } \right)}} \hfill \\
      \times {\rm{tr}}\left[ {{\Gamma _1}\left( {{\gamma _\tau }{g_{\delta \eta }} - {\gamma _\delta }{g_{\tau \eta }}} \right){\Gamma _2}\left( {\slashed{q} + {m_s}} \right){\gamma ^\tau }\left( {\slashed{k} + {m_s}} \right)} \right] \hfill \\ 
   \end{gathered}  \right\} \hfill \\
     \;\;\;\;\;\;\;\;\;\;\;\;\;\;\;\;\;\;\;\;\;\;\;\;\;\; + \frac{\partial }{{\partial {k_\eta }}}\int {{{\rm{d}}^4}q} \;\delta \left( {q - k} \right)\frac{\partial }{{\partial {q_\delta }}}\left\{ \begin{gathered}
     \;\;\;\frac{{{g^2}{{\left\langle {\bar qq} \right\rangle }^2}}}{{216\left( {{k^2} - m_{s}^2 + {\rm{i}}\varepsilon } \right)\left( {{q^2} - m_{s}^2 + {\rm{i}}\varepsilon } \right)}} \hfill \\
      \times {\rm{tr}}\left[ {{\Gamma _1}\left( {{\gamma _\tau }{g_{\delta \eta }} - {\gamma _\delta }{g_{\tau \eta }} - \frac{1}{2}{\rm{i}}{\sigma _{\delta \tau }}{\gamma _\eta }} \right){\Gamma _2}\left( {\slashed{q} + {m_s}} \right){\gamma ^\tau }\left( {\slashed{k} + {m_s}} \right)} \right] \hfill \\ 
   \end{gathered}  \right\}, \hfill \\
   \end{gathered} $
   \bigskip
   \par\noindent 
   $\begin{gathered}
     {\rm{Pack}_2}\left( {{\Gamma _1},{\Gamma _2}} \right) \equiv \int {{{\rm{d}}^4}q} \;\delta \left( {q - k} \right)\frac{\partial }{{\partial {q_\eta }}}\frac{\partial }{{\partial {q_\delta }}}\left\{ \begin{gathered}
     \;\;\;\frac{{{g^2}{{\left\langle {\bar ss} \right\rangle }^2}}}{{324\left( {{k^2} - m_q^2 + {\rm{i}}\varepsilon } \right)\left( {{q^2} - m_q^2 + {\rm{i}}\varepsilon } \right)}} \hfill \\
      \times {\rm{tr}}\left[ {{\Gamma _1}\left( {\slashed{k} + {m_q}} \right){\gamma ^\tau }\left( {\slashed{q} + {m_q}} \right){\Gamma _2}\left( {{\gamma _\tau }{g_{\delta \eta }} - {\gamma _\delta }{g_{\tau \eta }}} \right)} \right] \hfill \\ 
   \end{gathered}  \right\} \hfill \\
     \;\;\;\;\;\;\;\;\;\;\;\;\;\;\;\;\;\;\;\;\;\;\;\;\;\; + \frac{\partial }{{\partial {k_\eta }}}\int {{{\rm{d}}^4}q} \;\delta \left( {q - k} \right)\frac{\partial }{{\partial {q_\delta }}}\left\{ \begin{gathered}
     \;\;\;\frac{{{g^2}{{\left\langle {\bar ss} \right\rangle }^2}}}{{216\left( {{k^2} - m_q^2 + {\rm{i}}\varepsilon } \right)\left( {{q^2} - m_q^2 + {\rm{i}}\varepsilon } \right)}} \hfill \\
      \times {\rm{tr}}\left[ {{\Gamma _1}\left( {\slashed{k} + {m_q}} \right){\gamma ^\tau }\left( {\slashed{q} + {m_q}} \right){\Gamma _2}\left( { - 2{\gamma _\tau }{g_{\delta \eta }} + 2{\gamma _\delta }{g_{\tau \eta }} + \frac{1}{2}{\rm{i}}{\sigma _{\delta \tau }}{\gamma _\eta }} \right)} \right]\hfill \\ 
   \end{gathered}  \right\} \,. \hfill \\
   \end{gathered} $

\end{widetext}
\end{document}